\documentclass[10pt]{article}
\usepackage{graphicx}
\usepackage{amsmath}
\usepackage{amssymb}
\usepackage{caption2}
\begin{document}
\begin{center}
{\large {\bf \sc{  Possible assignments of the  $X(3872)$, $Z_c(3900)$ and $Z_b(10610)$ as axial-vector molecular states
  }}} \\[2mm]
Zhi-Gang  Wang$^{1}$ \footnote{E-mail: zgwang@aliyun.com.  }, Tao Huang$^{2}$ \footnote{Email: huangtao@ihep.ac.cn}     \\
$^{1}$ Department of Physics, North China Electric Power University, Baoding 071003, P. R. China \\
$^{2}$ Institute of High Energy Physics and Theoretical Physics
Center for Science Facilities, Chinese Academy of Sciences, Beijing 100049, P.R. China
\end{center}

\begin{abstract}
In this article, we  construct both the color singlet-singlet type and octet-octet type currents to interpolate the $X(3872)$, $Z_c(3900)$, $Z_b(10610)$,
and calculate the  vacuum condensates up to dimension-10  in the operator product expansion. Then we study  the axial-vector  hidden charmed and hidden bottom molecular states with the QCD sum rules, explore the energy scale dependence of the QCD sum rules for the heavy molecular states  in details,
and use the  formula $\mu=\sqrt{M^2_{X/Y/Z}-(2{\mathbb{M}}_Q)^2}$
  with the effective masses ${\mathbb{M}}_Q$ to determine the energy scales.  The numerical results  support assigning
  the $X(3872)$, $Z_c(3900)$, $Z_b(10610)$ as the color singlet-singlet type  molecular states  with $J^{PC}=1^{++}$, $1^{+-}$, $1^{+-}$, respectively,
  more theoretical and experimental works are still needed to distinguish the molecule and tetraquark assignments; while there are no
  candidates for the color octet-octet type molecular states.
\end{abstract}



\section{Introduction}

In 2003, the  Belle collaboration  reported the first observation of  the charmonium-like state $X(3872)$ in the $\pi^+ \pi^- J/\psi$ mass spectrum in the exclusive
 processes  $B^\pm \to K^\pm \pi^+ \pi^- J/\psi$  \cite{X3872-2003}.
   The evidences for the decay modes $X(3872) \to \gamma J/\psi, \, \gamma \psi^{\prime}$  imply the positive charge conjugation $C=+$ \cite{X3872-Jpsi-gamma}, while
    angular correlations between the final state particles in the $\pi^+ \pi^- J/\psi$ support the $J^{PC}=1^{++}$ assignment, and
    strongly disfavor (or exclude) the $0^{++}$, $0^{-+}$, $1^{-+}$, $2^{-+}$  assignments \cite{X3872-JPC}.
 The $X(3872)$ has been extensively studied since its first observation, for more articles on this subject, one can consult the reviews \cite{Swanson2006}.

 In 2011, the Belle collaboration reported the first observation of the $Z_b(10610)$ and $Z_b(10650)$ in the $\pi^{\pm}\Upsilon({\rm 1,2,3S})$  and $\pi^{\pm} h_b({\rm 1,2P})$   mass spectra in the exclusive  processes $\Upsilon({\rm 5S}) \to \Upsilon({\rm 1,2,3S})\, \pi^+ \pi^-$, $h_b({\rm 1,2P})\,\pi^+\pi^-$  \cite{Belle1105}. The quantum numbers (isospin, G-parity, spin and parity)  $I^G(J^P)=1^+(1^+)$ are favored \cite{Belle1105}.
Later, the Belle collaboration updated the parameters  $ M_{Z_b(10610)}=(10607.2\pm2.0)\,\rm{ MeV}$, $M_{Z_b(10650)}=(10652.2\pm1.5)\,\rm{MeV}$,
$\Gamma_{Z_b(10610)}=(18.4\pm2.4) \,\rm{MeV}$ and
$\Gamma_{Z_b(10650)}=(11.5\pm2.2)\,\rm{ MeV}$ \cite{Belle1110}.
In 2013, the Belle collaboration reported the first observation of the decay processes $\Upsilon(5{\rm S}) \to \Upsilon ({\rm 1,2,3S}) \,\pi^0 \pi^0$, and obtained  the neutral particle $Z_b^0(10610)$ in a Dalitz analysis of the decays to $\Upsilon(2,3{\rm S})\, \pi^0$ \cite{Belle1308}.
There have been several assignments of the  $Z_b(10610)$ and $Z_b(10650)$, such as the molecular states \cite{Molecule-Zb}, tetraquark states \cite{Tetraquark-Zb}, threshold cusps \cite{Cusp-Zb}, rescattering effects \cite{Rescatter-Zb}, etc.

 In 2013, the BESIII collaboration reported the first observation of the structure $Z_c(3900)$ in the $\pi^\pm J/\psi$ mass spectrum in
 the process  $e^+e^- \to \pi^+\pi^-J/\psi$ \cite{BES3900}. The  mass and decay width are  $(3899.0\pm 3.6\pm 4.9)\,\rm{ MeV}$ and $(46\pm 10\pm 20) \,\rm{MeV}$,
 respectively \cite{BES3900}. Then the $Z_c(3900)$ was confirmed by the Belle and CLEO collaborations \cite{Belle3900,CLEO3900}.
 There have been several assignments, such as the molecular state \cite{Molecular3900}, tetraquark state \cite{Tetraquark3900},
  hadro-charmonium \cite{hadro-charmonium-3900}, rescattering effect \cite{FSI3900}, etc.

In this article, we will focus on the scenario of molecular states. In Ref.\cite{Nielsen-0803}, S. H. Lee et al take the $X(3872)$ as the
$D^{\ast0}\bar{D}^{0}-D^0\bar{D}^{\ast0}$   molecular state with $J^{PC}=1^{++}$, study its mass with the QCD sum rules   by calculating
 the vacuum condensates up  to dimension-6 in the operator product expansion, and obtain the value $M_{X(3872)}=(3.88 \pm 0.06)\,\rm{ GeV}$.
In Ref.\cite{ZhangJR0906}, J. R. Zhang and M. Q. Huang study the masses of the $Q\bar{q}\bar{Q}q$ type  molecular states with  QCD sum rules in a systematic way
by  calculating  the vacuum condensates up  to dimension-6.
In Ref.\cite{Nielsen0907}, R. D. Matheus et al   take the  $X(3872)$ as a mixture between charmonium and exotic molecular state with $J^{PC}=1^{++}$,
study the mass $M_{X(3872)}$ and decay width $\Gamma_{X(3872)\to J/\psi \pi^+\pi^-}$ with the QCD sum rules, and
conclude that the $X(3872)$ is approximately $97\%$ a charmonium state $\bar{c}c$ and $3\%$ a  molecular state $D^*\bar{D}$.
In Ref.\cite{ZhangJR1105}, J. R. Zhang et al take the  $Z_b(10610)$ as a bottomonium-like molecular state $B^*\bar{B}$, study its mass with
the QCD sum rules   by calculating  the vacuum condensates up  to dimension-6, and obtain the value $M_{Z_b}=( 10.54\pm0.22)\,\rm{ GeV}$.
 In Ref.\cite{ChenW1305}, W. Chen et al  take the $X(3872)$ as the $J^{PC}=1^{++}$  mixed state of the charmonium hybrid and $D^*{\bar D}$ molecular state,
  study its mass with the QCD sum rules, and observe that the mixing is robust.
In Ref.\cite{ZhangJR1304}, J. R. Zhang takes the $Z_c(3900)$ as the $D^*{\bar D}$ molecular state without distinguishing  its charge conjugation,
study the mass with the QCD sum rules by calculating the vacuum condensates up to dimension-9, and obtain the value $M_{Z_c}=(3.86 \pm 0.27)\,\rm{ GeV}$.

In all those works \cite{Nielsen-0803,ZhangJR0906,Nielsen0907,ZhangJR1105,ChenW1305,ZhangJR1304}, the $\overline{MS}$ masses are taken, however,
the energy scales at which the QCD spectral densities are calculated are either not shown explicitly or not specified, and the energy scale dependence of
the QCD sum rules is not studied.
In the QCD sum rules for the hidden charmed (or bottom) tetraquark states and molecular states, the integrals
 \begin{eqnarray}
 \int_{4m_Q^2}^{s_0} ds \rho_{QCD}(s)\exp\left(-\frac{s}{T^2} \right)\, ,
 \end{eqnarray}
are sensitive to the heavy quark masses $m_Q$, where the $\rho_{QCD}(s)$ denotes the QCD spectral densities and the $T^2$ denotes
the Borel parameters. Variations of the heavy quark masses lead to changes of integral ranges $4m_Q^2-s_0$ of the variable
$\bf{ds}$ besides the QCD spectral densities, therefore changes of the Borel windows and predicted masses and pole residues. Furthermore,
in Refs.\cite{Nielsen-0803,ZhangJR0906,Nielsen0907,ZhangJR1105,ChenW1305,ZhangJR1304}, the higher dimensional  vacuum condensates are neglected
in one way or another.  The higher dimensional vacuum condensates play an important role in determining the Borel windows,
although they play a less important role in the Borel windows.

In Refs.\cite{WangHuangTao,Wang1311,Wang1312,WangHuangTao1312}, we focus on the scenario of tetraquark states, distinguish
the charge conjugations of the interpolating  currents, calculate the  vacuum condensates up to dimension-10  in
the operator product expansion, study the diquark-antidiquark type scalar, vector, axial-vector, tensor hidden charmed tetraquark states and
axial-vector hidden bottom tetraquark states systematically  with the QCD sum rules, make reasonable  assignments of the $X(3872)$,
$Z_c(3900)$, $Z_c(3885)$, $Z_c(4020)$, $Z_c(4025)$, $Z(4050)$, $Z(4250)$, $Y(4360)$, $Y(4630)$, $Y(4660)$, $Z_b(10610)$  and $Z_b(10650)$.
Furthermore,  we  explore the energy scale dependence of the QCD sum rules for the hidden charmed and hidden bottom tetraquark states
in details for the first time, and suggest a  formula,
\begin{eqnarray}
\mu&=&\sqrt{M^2_{X/Y/Z}-(2{\mathbb{M}}_Q)^2} \, ,
 \end{eqnarray}
 with the effective masses ${\mathbb{M}}_c=1.80\,\rm{GeV}$ and ${\mathbb{M}}_b=5.13\,\rm{GeV}$ to determine the energy scales of the  QCD spectral densities,
 which works  well.

 In this article, we take the $X(3872)$, $Z_c(3900)$, $Z_b(10610)$ as the axial-vector  hadronic molecular states, distinguish
the charge conjugations, construct  both the color singlet-singlet type   currents   and color octet-octet type currents to interpolate  them.
We calculate the contributions of the vacuum condensates up to dimension-10, study the masses and pole residues,
and explore the energy scale dependence in details so as to see whether or not the formula $\mu=\sqrt{M^2_{X/Y/Z}-(2{\mathbb{M}}_Q)^2}$ survives
 in the case of the molecular states, and make tentative assignments  of the $X(3872)$, $Z_c(3900)$, $Z_b(10610)$ in the scenario of molecular states.

The article is arranged as follows:  we derive the QCD sum rules for
the masses and pole residues of  the axial-vector molecular states  in section 2; in section 3,
we present the numerical results and discussions; section 4 is reserved for our conclusion.

\section{QCD sum rules for  the   $J^{P}=1^{+}$ molecular  states }
In the following, we write down  the two-point correlation functions $\Pi_{\mu\nu}(p)$  in the QCD sum rules,
\begin{eqnarray}
\Pi_{\mu\nu}(p)&=&i\int d^4x e^{ip \cdot x} \langle0|T\left\{J_\mu(x)J_\nu^{\dagger}(0)\right\}|0\rangle \, , \\
J^0_\mu(x)&=&\frac{\bar{u}(x)i\gamma_5 Q(x)\bar{Q}(x)\gamma_\mu d(x)+t\bar{u}(x)\gamma_\mu  Q(x)\bar{Q}(x)i\gamma_5 d(x)}{\sqrt{2}}  \, ,  \\
J^8_\mu(x)&=&\frac{\bar{u}(x)i\gamma_5 \lambda^a Q(x)\bar{Q}(x)\gamma_\mu \lambda^a d(x)+t\bar{u}(x)\gamma_\mu \lambda^a Q(x)\bar{Q}(x)i\gamma_5 \lambda^a d(x)}{\sqrt{2}} \, ,
\end{eqnarray}
where $t=\pm 1$, $J_\mu(x)=J^0_{\mu}(x),\,J^8_{\mu}(x)$, the $\lambda^a$ is the Gell-Mann matrix. We construct the color singlet-singlet type (0-0 type) currents $J^0_\mu(x)$ (see Refs.\cite{Nielsen-0803,ZhangJR0906,Nielsen0907,ZhangJR1105,ChenW1305,ZhangJR1304}) and color octet-octet type (8-8 type) currents $J^8_\mu(x)$ (see Ref.\cite{Wang-NPA}) to study the hadronic molecular states $X(3872)$ (to be more precise, the charged partner of the $X(3872)$), $Z_c(3900)$, $Z_b(10610)$, etc.
We can rearrange the 8-8 type currents $J^8_\mu(x)$ in terms of the following 0-0 type currents,
 \begin{eqnarray}
 J^8_\mu(x)&=&\frac{\sqrt{2}}{4}\bar{u}(x)i\gamma_5d(x)\bar{Q}(x)\gamma_\mu Q(x)+t\frac{\sqrt{2}}{4}\bar{Q}(x)i\gamma_5Q(x)\bar{u}(x)\gamma_\mu d(x) \nonumber\\
 &&+\frac{\sqrt{2}i}{4}\bar{u}(x)\gamma_5\gamma_\alpha d(x)\bar{Q}(x)\gamma_\mu \gamma^\alpha Q(x)+t\frac{\sqrt{2}i}{4}\bar{Q}(x)\gamma_5\gamma_\alpha Q(x)\bar{u}(x)\gamma_\mu \gamma^\alpha d(x) \nonumber\\
 &&+\frac{\sqrt{2}i}{8}\bar{u}(x)\gamma_5\sigma_{\alpha\beta} d(x)\bar{Q}(x)\gamma_\mu \sigma^{\alpha\beta} Q(x)+t\frac{\sqrt{2}i}{8}\bar{Q}(x)\gamma_5\sigma_{\alpha\beta} Q(x)\bar{u}(x)\gamma_\mu \sigma^{\alpha\beta} d(x) \nonumber\\
 &&+\frac{\sqrt{2}i}{4}\bar{u}(x) \gamma_\alpha d(x)\bar{Q}(x)\gamma_\mu \gamma^\alpha\gamma_5 Q(x)+t\frac{\sqrt{2}i}{4}\bar{Q}(x) \gamma_\alpha Q(x)\bar{u}(x)\gamma_\mu \gamma^\alpha\gamma_5 d(x) \nonumber\\
 &&+\frac{\sqrt{2}i}{4}\bar{u}(x)d(x)\bar{Q}(x)\gamma_\mu \gamma_5 Q(x)+t\frac{\sqrt{2}i}{4}\bar{Q}(x)Q(x)\bar{u}(x)\gamma_\mu \gamma_5 d(x) \nonumber\\
 &&-\frac{\sqrt{2}}{3}\bar{u}(x)i\gamma_5 Q(x)\bar{Q}(x)\gamma_\mu d(x)-t\frac{\sqrt{2}}{3}\bar{u}(x)\gamma_\mu  Q(x)\bar{Q}(x)i\gamma_5 d(x) \, ,
 \end{eqnarray}
 with the identity,
 \begin{eqnarray}
 \lambda^a_{ij}\lambda^a_{mn}&=&2\delta_{in}\delta_{mj}-\frac{2}{3}\delta_{ij}\delta_{mn}\, ,
 \end{eqnarray}
 in the color space. The 8-8 type current can be taken as a special  superposition of the 0-0 type currents.
 Under charge conjugation transform $\widehat{C}$, the currents $J_\mu(x)$ have the properties,
\begin{eqnarray}
\widehat{C}J_{\mu}(x)\widehat{C}^{-1}&=&\mp J_\mu(x)\mid_{u {\leftrightarrow}d} \,\,\,\, {\rm for}\,\,\,\, t=\pm1\, .
\end{eqnarray}
The values $t=\mp 1$ correspond to  the positive and negative charge conjugations, respectively.

We can insert  a complete set of intermediate hadronic states with
the same quantum numbers as the current operators $J_\mu(x)$ into the
correlation functions $\Pi_{\mu\nu}(p)$  to obtain the hadronic representation
\cite{SVZ79,Reinders85}. After isolating the ground state
contributions from the pole terms,  we get the following results,
\begin{eqnarray}
\Pi_{\mu\nu}(p)&=&\frac{\lambda_{X/Z}^2}{M^2_{X/Z}-p^2}\left(-g_{\mu\nu} +\frac{p_\mu p_\nu}{p^2}\right) +\cdots =\Pi(p)\left(-g_{\mu\nu} +\frac{p_\mu p_\nu}{p^2}\right) +\cdots\, \, ,
\end{eqnarray}
where the pole residues (or couplings) $\lambda_{X/Z}$ are defined by
\begin{eqnarray}
 \langle 0|J_\mu(0)|X/Z(p)\rangle=\lambda_{X/Z}\, \varepsilon_\mu \, ,
\end{eqnarray}
the $\varepsilon_\mu$ are the polarization vectors of the axial-vector mesons $X(3872)$, $Z_c(3900)$, $Z_b(10610)$, etc.

Here we take a short digression to discus the possible contaminations originate from the higher resonances and continuum states.
In the following, we will discus the hidden-charmed systems for simplicity, the conclusion survives in the hidden-bottom systems.
In the nonrelativistic and heavy quark limit,
the $C=+$ currents are reduced to the  forms,
\begin{eqnarray}
\bar{u}\gamma^5 c \,\bar{c}\gamma^j d-\bar{u}\gamma^j c \,\bar{c}\gamma^5 d&\propto&\xi^\dagger_u  \xi_c \, \xi_c^\dagger \frac{\sigma^j}{2} \xi_d
-\xi^\dagger_u \frac{\sigma^j}{2} \xi_c \, \xi_c^\dagger  \xi_d   \, , \nonumber\\
\bar{u}\gamma^j c \,\bar{c}\gamma^k d+\bar{u}\gamma^k c \,\bar{c}\gamma^j d&\propto& \xi^\dagger_u \frac{\sigma^j}{2} \xi_c \, \xi_c^\dagger \frac{\sigma^k}{2} \xi_d
+\xi^\dagger_u \frac{\sigma^k}{2} \xi_c \, \xi_c^\dagger \frac{\sigma^j}{2} \xi_d\, ,
\end{eqnarray}
while the $C=-$ currents are reduced to the forms,
\begin{eqnarray}
\bar{u}\gamma^5 c \,\bar{c}\gamma^j d+\bar{u}\gamma^j c \,\bar{c}\gamma^5 d&\propto&\xi^\dagger_u  \xi_c \, \xi_c^\dagger \frac{\sigma^j}{2} \xi_d
+\xi^\dagger_u \frac{\sigma^j}{2} \xi_c \, \xi_c^\dagger  \xi_d\, , \nonumber\\
\bar{u}\gamma^j c \,\bar{c}\gamma^k d-\bar{u}\gamma^k c \,\bar{c}\gamma^j d&\propto& \xi^\dagger_u \frac{\sigma^j}{2} \xi_c \, \xi_c^\dagger \frac{\sigma^k}{2} \xi_d
-\xi^\dagger_u \frac{\sigma^k}{2} \xi_c \, \xi_c^\dagger \frac{\sigma^j}{2} \xi_d\, ,
\end{eqnarray}
where the $\xi_{c,u,d}$ are the two-component quark fields and   the $\sigma^i$ are the pauli matrixes. The bilinear fields $\xi^\dagger_i \xi_j$ and
 $\xi^\dagger_i \frac{\sigma^k}{2} \xi_j$ have the spins 0 and 1, respectively, and couple  to (pseudo-) scalar and (axial-) vector meson fields, respectively.
The currents $J^0_\mu$ with $C=\pm$ couple potentially to the $\frac{D\bar{D}^*{\mp} D^*\bar{D}}{\sqrt{2}}$ molecular or scattering  states, while the currents
$J^0_{\mu\nu}=\bar{u}\gamma_\mu  c \,\bar{c}\gamma_\nu  d\pm\bar{u}\gamma_\nu  c \,\bar{c}\gamma_\mu  d$ with $C=\pm$ couple potentially
to the $D^* \bar{D}^*$ molecular or scattering states.

On the other hand, the octet currents are reduced into the following forms,
\begin{eqnarray}
\bar{u}i\gamma^5 \lambda^a c \, \bar{c}\lambda^a\gamma^j  d&\propto& \xi^\dagger_u \lambda^a \xi_c \, \xi_c^\dagger  \lambda^a \frac{\sigma^j}{2}\xi_d=
4\xi^\dagger_u \frac{\sigma^k}{2}\frac{\sigma^j}{2} \xi_d\, \xi^\dagger_c \frac{\sigma^k}{2}\xi_c+\xi^\dagger_u \frac{\sigma^j}{2}\xi_d \, \xi^\dagger_c \xi_c
 -\frac{2}{3}\xi^\dagger_u  \xi_c\, \xi_c^\dagger  \frac{\sigma^j}{2} \xi_d  \nonumber\\
&=&\xi^\dagger_u  \xi_d\, \xi^\dagger_c \frac{\sigma^j}{2}\xi_c-2\epsilon^{ijk}\xi^\dagger_u \frac{\sigma^i}{2} \xi_d\, \xi^\dagger_c \frac{\sigma^k}{2}\xi_c
+\xi^\dagger_u \frac{\sigma^j}{2}\xi_d\, \xi^\dagger_c \xi_c -\frac{2}{3}\xi^\dagger_u  \xi_c \,\xi_c^\dagger  \frac{\sigma^j}{2} \xi_d \, , \nonumber\\
\bar{u}\lambda^a\gamma^j  c \,\bar{c}\lambda^a\gamma^k  d&\propto& \xi^\dagger_u \lambda^a\frac{\sigma^j}{2} \xi_c \,\xi_c^\dagger \lambda^a\frac{\sigma^k}{2} \xi_d\nonumber\\
&=& 4\xi^\dagger_u\frac{\sigma^i}{2}\frac{\sigma^k}{2} \xi_d \, \xi_c^\dagger\frac{\sigma^i}{2} \frac{\sigma^j}{2} \xi_c +\xi^\dagger_u\frac{\sigma^k}{2} \xi_d \, \xi_c^\dagger \frac{\sigma^j}{2} \xi_c -\frac{2}{3}\xi^\dagger_u
\frac{\sigma^j}{2} \xi_c \, \xi_c^\dagger \frac{\sigma^k}{2} \xi_d \nonumber\\
&=& \frac{\delta_{jk}}{4}\xi^\dagger_u \xi_d \, \xi_c^\dagger \xi_c
+\frac{1}{2}\epsilon^{jkm} \xi^\dagger_u\frac{\sigma^m}{2} \xi_d \, \xi_c^\dagger \xi_c
+\frac{1}{2}\epsilon^{kjm} \xi^\dagger_u\xi_d \, \xi_c^\dagger \frac{\sigma^m}{2} \xi_c
 \nonumber\\
&&+\epsilon^{ikm}\epsilon^{ijn} \xi^\dagger_u\frac{\sigma^m}{2} \xi_d \, \xi_c^\dagger \frac{\sigma^n}{2} \xi_c+\xi^\dagger_u\frac{\sigma^k}{2} \xi_d \, \xi_c^\dagger \frac{\sigma^j}{2} \xi_c -\frac{2}{3}\xi^\dagger_u
\frac{\sigma^j}{2} \xi_c \,\xi_c^\dagger \frac{\sigma^k}{2} \xi_d \, .
\end{eqnarray}
The octet current $J^8_\mu=\bar{u}i\gamma^5 \lambda^a c \, \bar{c}\lambda^a\gamma_\mu  d$ couples potentially to the $J/\psi\pi$, $\psi(3770)\pi$, $\eta_c \rho$, $J/\psi\rho$, $D\bar{D}^*$ molecular or scattering states.
The octet current $J^8_{\mu\nu}=\bar{u}\lambda^a\gamma_\mu  c \,\bar{c}\lambda^a\gamma_\nu  d$ couples potentially to the $\eta_c \pi$,  $\eta_c \rho$, $J/\psi\pi$, $\psi(3770)\pi$, $J/\psi \rho$, $D^*\bar{D}^*$ molecular or scattering states.
In this article, we take the currents $J^{0,8}_{\mu}$ not the currents $J^{0,8}_{\mu\nu}$, the $D^*\bar{D}^*$ molecular or scattering states  have no contaminations.

In the scenario of meta-stable Feshbach resonances, the $X(3872)$, $Z_c(3900)$, $Z_c(4025)$, $Z_b(10610)$, $Z_b(10650)$ are taken as the
$J/\psi\rho-D\bar{D}^*$, $\psi(3770)\pi-D\bar{D}^*$, $h_c({\rm 2P})\pi-D^*\bar{D}^*$, $\chi_{b0}\rho-B\bar{B}^*$, $\chi_{b1}\rho-B^*\bar{B}^*$ hadrocharmonium-molecule
 mixed states, respectively, where the $\chi_{b0}\rho$ and $\chi_{b1}\rho$ are P-wave systems \cite{Italy1311}. The hadrocharmonium system admits bound states giving rise to a
discrete spectrum of levels, a resonance occurs if one of such levels falls close to some
open-charm threshold, as  the coupling between channels leads  to an attractive interaction and favors the formation of a meta-stable Feshbach
resonance. The couplings of the currents $J_\mu$ to the near-threshold  hadrocharmonium states $J/\psi\rho$, $\psi(3770)\pi$ and  $\chi_{b0}\rho$ contribute to the molecular states $X(3872)$, $Z_c(3900)$ and
 $Z_b(10610)$, respectively.

   Now we study the contributions of the  intermediate   meson-loops (or the scattering states  $D
D^\ast$, $J/\psi \pi$, $J/\psi \rho$, etc) to the correlation functions $\Pi_{\mu\nu}(p)$,
\begin{eqnarray}
\Pi_{\mu\nu}(p)&=&-\frac{\widehat{\lambda}_{X/Z}^{2}}{ p^2-\widehat{M}_{X/Z}^2}\widetilde{g}_{\mu\nu}(p)-\frac{\widehat{\lambda}_{X/Z}}{p^2-\widehat{M}_{X/Z}^2}\widetilde{g}_{\mu\alpha}(p)
 \Sigma_{DD^*}(p) \widetilde{g}^{\alpha\beta}(p) \widetilde{g}_{\beta\nu}(p)\frac{\widehat{\lambda}_{X/Z}}{p^2-\widehat{M}_{X/Z}^2} \nonumber \\
 &&-\frac{\widehat{\lambda}_{X/Z}}{p^2-\widehat{M}_{X/Z}^2}\widetilde{g}_{\mu\alpha}(p)
 \Sigma_{J/\psi\pi}(p) \widetilde{g}^{\alpha\beta}(p) \widetilde{g}_{\beta\nu}(p)\frac{\widehat{\lambda}_{X/Z}}{p^2-\widehat{M}_{X/Z}^2} \nonumber \\
 &&-\frac{\widehat{\lambda}_{X/Z}}{p^2-\widehat{M}_{X/Z}^2}\widetilde{g}_{\mu\alpha}(p)
 \Sigma_{J/\psi\rho}^{\alpha\beta}(p)   \widetilde{g}_{\beta\nu}(p)\frac{\widehat{\lambda}_{X/Z}}{p^2-\widehat{M}_{X/Z}^2} +\cdots \, ,\nonumber \\
 &=&-\frac{\widehat{\lambda}_{X/Z}^{2}}{ p^2-\widehat{M}_{X/Z}^2-\Sigma_{DD^*}(p)-\Sigma_{J/\psi\pi}(p)-\Sigma_{J/\psi\rho}(p)+\cdots}\widetilde{g}_{\mu\nu}(p)+\cdots \, , \end{eqnarray}
where
\begin{eqnarray}
\Sigma_{DD^*}(p)&=&i\int~{d^4q\over(2\pi)^4}\frac{G^2_{X/ZDD^*}}{\left[q^2-M_{D}^2\right]\left[ (p-q)^2-M_{D^*}^2\right]} \, ,\\
\Sigma_{J/\psi \pi}(p)&=&i\int~{d^4q\over(2\pi)^4}\frac{G^2_{X/ZJ/\psi \pi}}{\left[q^2-M_{J/\psi}^2\right]\left[ (p-q)^2-M_{\pi}^2\right]} \, ,\\
\Sigma_{J/\psi\rho}^{\alpha\beta}(p)&=&i\int~{d^4q\over(2\pi)^4}\frac{G^2_{X/ZJ/\psi \rho}\epsilon^{\alpha\theta\sigma\tau}\epsilon^{\beta\theta^{\prime}\sigma^{\prime}\tau^{\prime}}p_\tau p_{\tau^\prime}\widetilde{g}_{\theta\theta^{\prime}}(q)\widetilde{g}_{\sigma\sigma^{\prime}}(p-q)}{\left[
q^2-M_{J/\psi}^2\right]\left[ (p-q)^2-M_{\rho}^2\right]}\, , \nonumber\\
&=&\Sigma_{J/\psi \rho}(p)\widetilde{g}^{\alpha\beta}(p)+\Sigma_{J/\psi \rho}^1(p) \frac{p^{\alpha}p^{\beta} }{p^2}\, ,
\end{eqnarray}
$\widetilde{g}_{\mu\nu}(p)=-g_{\mu\nu}+\frac{p_{\mu}p_{\nu}}{p^2}$, the $G_{X/Z DD^*}$, $G_{X/Z J/\psi\pi}$, $G_{X/Z J/\psi\rho}$ are
hadronic coupling constants, the $\widehat{\lambda}_{X/Z}$ and $\widehat{M}_{X/Z}$ are bare quantities to absorb the divergences
in the self-energies $\Sigma_{DD^*}(p)$, $\Sigma_{J/\psi \pi}(p)$, $\Sigma_{J/\psi \rho}(p)$, etc.

The renormalized self-energies  contribute  a finite imaginary part to modify the dispersion relation,
\begin{eqnarray}
\Pi_{\mu\nu}(p) &=&-\frac{\lambda_{X/Z}^{2}}{ p^2-M_{X/Z}^2+i\sqrt{p^2}\Gamma(p^2)}\widetilde{g}_{\mu\nu}(p)+\cdots \, ,
 \end{eqnarray}
the physical  widths $\Gamma_{Z_c(3900)}(M_Z^2)=(46 \pm 10 \pm 20)\, \rm{MeV}$ and $\Gamma_{X(3872)}(M_X^2)<1.2\,\rm{MeV}$ are small enough,
 the zero width approximation in  the hadronic spectral densities works. The discussion survives  in the hidden-bottom systems according to
  the small physical widths $\Gamma_{Z_b(10610)}=(18.4\pm2.4) \,\rm{MeV}$ and $\Gamma_{Z_b(10650)}=(11.5\pm2.2)\,\rm{ MeV}$.
  The contaminations of the intermediate meson-loops are expected
 to be small.

 In the following,  we briefly outline  the operator product expansion for the correlation functions $\Pi_{\mu\nu}(p)$  in perturbative
QCD.  We contract the quark fields in the correlation functions
$\Pi_{\mu\nu}(p)$ with Wick theorem, obtain the results:
\begin{eqnarray}
\Pi_{\mu\nu}^{0}(p)&=&-\frac{i}{2} \delta_{jk}\delta_{mn}\delta_{k^{\prime}j^{\prime}}\delta_{n^{\prime}m^{\prime}}   \int d^4x e^{ip \cdot x}   \nonumber\\
&&\left\{{\rm Tr}\left[ \gamma_5 S_Q^{kk^{\prime}}(x)\gamma_5 S^{j^{\prime}j}(-x)\right] {\rm Tr}\left[ \gamma_\mu S^{nn^{\prime}}(x)\gamma_\nu S_Q^{m^{\prime}m}(-x)\right] \right. \nonumber\\
&&+{\rm Tr}\left[ \gamma_\mu S_Q^{kk^{\prime}}(x)\gamma_\nu S^{j^{\prime}j}(-x)\right] {\rm Tr}\left[ \gamma_5 S^{nn^{\prime}}(x)\gamma_5 S_Q^{m^{\prime}m}(-x)\right] \nonumber\\
&&\pm {\rm Tr}\left[ \gamma_\mu S_Q^{kk^{\prime}}(x)\gamma_5 S^{j^{\prime}j}(-x)\right] {\rm Tr}\left[ \gamma_5 S^{nn^{\prime}}(x)\gamma_\nu S_Q^{m^{\prime}m}(-x)\right] \nonumber\\
 &&\left.\pm {\rm Tr}\left[ \gamma_5 S_Q^{kk^{\prime}}(x)\gamma_\nu S^{j^{\prime}j}(-x)\right] {\rm Tr}\left[ \gamma_\mu S^{nn^{\prime}}(x)\gamma_5 S_Q^{m^{\prime}m}(-x)\right] \right\} \, , \\
 \Pi_{\mu\nu}^{8}(p)&=&\Pi_{\mu\nu}^{0}(p)\mid_{\delta_{jk}\delta_{mn}\delta_{k^{\prime}j^{\prime}}\delta_{n^{\prime}m^{\prime}} \to \lambda^a_{jk}\lambda^a_{mn}\lambda^b_{k^{\prime}j^{\prime}}\lambda^b_{n^{\prime}m^{\prime}}}
\end{eqnarray}
where the $\mp$ correspond  the positive and negative charge conjugations, respectively,
 the $S^{ij}(x)$  and $S^{ij}_Q(x)$ are the full light  and heavy quark propagators, respectively,
\begin{eqnarray}
S^{ij}(x)&=& \frac{i\delta_{ij}\!\not\!{x}}{ 2\pi^2x^4}-\frac{\delta_{ij}\langle
\bar{q}q\rangle}{12} -\frac{\delta_{ij}x^2\langle \bar{q}g_s\sigma Gq\rangle}{192} -\frac{ig_sG^{a}_{\alpha\beta}t^a_{ij}(\!\not\!{x}
\sigma^{\alpha\beta}+\sigma^{\alpha\beta} \!\not\!{x})}{32\pi^2x^2} -\frac{i\delta_{ij}x^2\!\not\!{x}g_s^2\langle \bar{q} q\rangle^2}{7776}\nonumber\\
&&  -\frac{\delta_{ij}x^4\langle \bar{q}q \rangle\langle g_s^2 GG\rangle}{27648} -\frac{1}{8}\langle\bar{q}_j\sigma^{\mu\nu}q_i \rangle \sigma_{\mu\nu}-\frac{1}{4}\langle\bar{q}_j\gamma^{\mu}q_i\rangle \gamma_{\mu }+\cdots \, ,
\end{eqnarray}
\begin{eqnarray}
S^{ij}_Q(x)&=&\frac{i}{(2\pi)^4}\int d^4k e^{-ik \cdot x} \left\{
\frac{\delta_{ij}}{\!\not\!{k}-m_Q}
-\frac{g_sG^n_{\alpha\beta}t^n_{ij}}{4}\frac{\sigma^{\alpha\beta}(\!\not\!{k}+m_Q)+(\!\not\!{k}+m_Q)
\sigma^{\alpha\beta}}{(k^2-m_Q^2)^2}\right.\nonumber\\
&&\left. +\frac{g_s D_\alpha G^n_{\beta\lambda}t^n_{ij}(f^{\lambda\beta\alpha}+f^{\lambda\alpha\beta}) }{3(k^2-m_Q^2)^4}-\frac{g_s^2 (t^at^b)_{ij} G^a_{\alpha\beta}G^b_{\mu\nu}(f^{\alpha\beta\mu\nu}+f^{\alpha\mu\beta\nu}+f^{\alpha\mu\nu\beta}) }{4(k^2-m_Q^2)^5}+\cdots\right\} \, ,\nonumber\\
f^{\lambda\alpha\beta}&=&(\!\not\!{k}+m_Q)\gamma^\lambda(\!\not\!{k}+m_Q)\gamma^\alpha(\!\not\!{k}+m_Q)\gamma^\beta(\!\not\!{k}+m_Q)\, ,\nonumber\\
f^{\alpha\beta\mu\nu}&=&(\!\not\!{k}+m_Q)\gamma^\alpha(\!\not\!{k}+m_Q)\gamma^\beta(\!\not\!{k}+m_Q)\gamma^\mu(\!\not\!{k}+m_Q)\gamma^\nu(\!\not\!{k}+m_Q)\, ,
\end{eqnarray}
and  $t^n=\frac{\lambda^n}{2}$,  $D_\alpha=\partial_\alpha-ig_sG^n_\alpha t^n$ \cite{Reinders85}, then compute  the integrals both in the coordinate and momentum spaces,  and obtain the correlation functions $\Pi_{\mu\nu}(p)$ therefore the spectral densities at the level of   quark-gluon degrees  of freedom.
In Eq.(21), we retain the terms $\langle\bar{q}_j\sigma_{\mu\nu}q_i \rangle$ and $\langle\bar{q}_j\gamma_{\mu}q_i\rangle$ originate from the Fierz re-ordering
of the $\langle q_i \bar{q}_j\rangle$ to  absorb the gluons  emitted from the heavy quark lines to form
$\langle\bar{q}_j g_s G^a_{\alpha\beta} t^a_{mn}\sigma_{\mu\nu} q_i \rangle$ and $\langle\bar{q}_j\gamma_{\mu}q_ig_s D_\nu G^a_{\alpha\beta}t^a_{mn}\rangle$ so
as to extract the mixed condensate and four-quark condensates $\langle\bar{q}g_s\sigma G q\rangle$ and $g_s^2\langle\bar{q}q\rangle^2$, respectively.

 Once analytical results are obtained,  we can take the
quark-hadron duality and perform Borel transform  with respect to
the variable $P^2=-p^2$ to obtain  the following QCD sum rules:
\begin{eqnarray}
\lambda^2_{X/Z}\, \exp\left(-\frac{M^2_{X/Z}}{T^2}\right)= \int_{4m_Q^2}^{s_0} ds\, \rho^{0/8}(s) \, \exp\left(-\frac{s}{T^2}\right) \, ,
\end{eqnarray}
where
\begin{eqnarray}
\rho^{0/8}(s)&=&\rho_{0}^{0/8}(s)+\rho_{3}^{0/8}(s) +\rho_{4}^{0/8}(s)+\rho_{5}^{0/8}(s)+\rho_{6}^{0/8}(s)+\rho_{7}^{0/8}(s) +\rho_{8}^{0/8}(s)+\rho_{10}^{0/8}(s)\, ,
\end{eqnarray}

\begin{eqnarray}
\rho_{0}^{0}(s)&=&\frac{1}{4096\pi^6}\int_{y_i}^{y_f}dy \int_{z_i}^{1-y}dz \, yz\,(1-y-z)^3\left(s-\overline{m}_Q^2\right)^2\left(35s^2-26s\overline{m}_Q^2+3\overline{m}_Q^4 \right)\, ,
\end{eqnarray}

\begin{eqnarray}
\rho_{0}^{8}(s)&=&\frac{1}{1152\pi^6}\int_{y_i}^{y_f}dy \int_{z_i}^{1-y}dz \, yz\,(1-y-z)^3\left(s-\overline{m}_Q^2\right)^2\left(35s^2-26s\overline{m}_Q^2+3\overline{m}_Q^4 \right)\, ,
\end{eqnarray}

\begin{eqnarray}
\rho_{3}^{0}(s)&=&-\frac{3m_Q\langle \bar{q}q\rangle}{256\pi^4}\int_{y_i}^{y_f}dy \int_{z_i}^{1-y}dz \, (y+z)(1-y-z)\left(s-\overline{m}_Q^2\right)\left(7s-3\overline{m}_Q^2 \right) \, ,
\end{eqnarray}

\begin{eqnarray}
\rho_{3}^{8}(s)&=&-\frac{m_Q\langle \bar{q}q\rangle}{24\pi^4}\int_{y_i}^{y_f}dy \int_{z_i}^{1-y}dz \, (y+z)(1-y-z)\left(s-\overline{m}_Q^2\right)\left(7s-3\overline{m}_Q^2 \right) \, ,
\end{eqnarray}

\begin{eqnarray}
\rho_{4}^{0}(s)&=&-\frac{m_Q^2}{3072\pi^4} \langle\frac{\alpha_s GG}{\pi}\rangle\int_{y_i}^{y_f}dy \int_{z_i}^{1-y}dz \left( \frac{z}{y^2}+\frac{y}{z^2}\right)(1-y-z)^3 \left\{ 8s-3\overline{m}_Q^2+\overline{m}_Q^4\, \delta\left(s-\overline{m}_Q^2\right)\right\} \nonumber\\
&&+\frac{1}{1024\pi^4}\langle\frac{\alpha_s GG}{\pi}\rangle\int_{y_i}^{y_f}dy \int_{z_i}^{1-y}dz\, (y+z)(1-y-z)^2 \,s\,(5s-4\overline{m}_Q^2) \, ,
\end{eqnarray}

\begin{eqnarray}
\rho_{4}^{8}(s)&=&-\frac{m_Q^2}{864\pi^4} \langle\frac{\alpha_s GG}{\pi}\rangle\int_{y_i}^{y_f}dy \int_{z_i}^{1-y}dz \left( \frac{z}{y^2}+\frac{y}{z^2}\right)(1-y-z)^3 \left\{ 8s-3\overline{m}_Q^2+\overline{m}_Q^4\, \delta\left(s-\overline{m}_Q^2\right)\right\} \nonumber\\
&&-\frac{1}{2304\pi^4}\langle\frac{\alpha_s GG}{\pi}\rangle\int_{y_i}^{y_f}dy \int_{z_i}^{1-y}dz\, (y+z)(1-y-z)^2 \,s\,(5s-4\overline{m}_Q^2) \nonumber\\
&&+t\frac{m_Q^2}{1152\pi^4}\langle\frac{\alpha_s GG}{\pi}\rangle\int_{y_i}^{y_f}dy \int_{z_i}^{1-y}dz \left(s-\overline{m}_Q^2 \right)\left\{ 7-2\left( \frac{1}{y}+ \frac{1}{z}\right) (1-y-z) \right. \nonumber\\
&&\left.+ \frac{7(1-y-z)^2}{2yz}  -\frac{7(1-y-z)}{2} +\left(\frac{1}{y}+\frac{1}{z} \right)\frac{(1-y-z)^2}{2}
 -\frac{7(1-y-z)^3}{12yz}  \right\}\, ,
\end{eqnarray}

\begin{eqnarray}
\rho_{5}^{0}(s)&=&\frac{3m_Q\langle \bar{q}g_s\sigma Gq\rangle}{512\pi^4}\int_{y_i}^{y_f}dy \int_{z_i}^{1-y}dz \,  (y+z) \left(5s-3\overline{m}_Q^2 \right) \nonumber\\
&&-\frac{3m_Q\langle \bar{q}g_s\sigma Gq\rangle}{256\pi^4}\int_{y_i}^{y_f}dy \int_{z_i}^{1-y}dz   \left(\frac{y}{z}+\frac{z}{y} \right)(1-y-z) \left(2s-\overline{m}_Q^2 \right)   \, ,
\end{eqnarray}

\begin{eqnarray}
\rho_{5}^{8}(s)&=&\frac{m_Q\langle \bar{q}g_s\sigma Gq\rangle}{48\pi^4}\int_{y_i}^{y_f}dy \int_{z_i}^{1-y}dz \,  (y+z) \left(5s-3\overline{m}_Q^2 \right) \nonumber\\
&&+\frac{m_Q\langle \bar{q}g_s\sigma Gq\rangle}{192\pi^4}\int_{y_i}^{y_f}dy \int_{z_i}^{1-y}dz   \left(\frac{y}{z}+\frac{z}{y} \right)(1-y-z) \left(2s-\overline{m}_Q^2 \right)  \nonumber\\
&&+t\frac{m_Q\langle \bar{q}g_s\sigma Gq\rangle}{576\pi^4}\int_{y_i}^{y_f}dy \int_{z_i}^{1-y}dz   \left(\frac{y}{z}+\frac{z}{y} \right)(1-y-z) \left(5s-3\overline{m}_Q^2 \right) \, ,
\end{eqnarray}

\begin{eqnarray}
\rho_{6}^{0}(s)&=&\frac{m_Q^2\langle\bar{q}q\rangle^2}{16\pi^2}\int_{y_i}^{y_f}dy +\frac{g_s^2\langle\bar{q}q\rangle^2}{864\pi^4}\int_{y_i}^{y_f}dy \int_{z_i}^{1-y}dz\, yz \left\{8s-3\overline{m}_Q^2 +\overline{m}_Q^4\, \delta\left(s-\overline{m}_Q^2 \right)\right\}\nonumber\\
&&-\frac{g_s^2\langle\bar{q}q\rangle^2}{576\pi^4}\int_{y_i}^{y_f}dy \int_{z_i}^{1-y}dz\, (1-y-z)\left\{ \left(\frac{z}{y}+\frac{y}{z} \right)\left(7s-4\overline{m}_Q^2 \right)\right.\nonumber\\
&&\left.+\frac{1}{3}\left(\frac{z}{y^2}+\frac{y}{z^2} \right)m_Q^2\left[ 7+5\overline{m}_Q^2\,\delta\left(s-\overline{m}_Q^2 \right)\right]-\frac{1}{3}(y+z)\left(4s-3\overline{m}_Q^2 \right)\right\} \, ,
\end{eqnarray}

\begin{eqnarray}
\rho_{6}^{8}(s)&=&\frac{2m_Q^2\langle\bar{q}q\rangle^2}{9\pi^2}\int_{y_i}^{y_f}dy +\frac{g_s^2\langle\bar{q}q\rangle^2}{243\pi^4}\int_{y_i}^{y_f}dy \int_{z_i}^{1-y}dz\, yz \left\{8s-3\overline{m}_Q^2 +\overline{m}_Q^4\, \delta\left(s-\overline{m}_Q^2 \right)\right\}\nonumber\\
&&+\frac{g_s^2\langle\bar{q}q\rangle^2}{1296\pi^4}\int_{y_i}^{y_f}dy \int_{z_i}^{1-y}dz\, (1-y-z)\left\{ \left(\frac{z}{y}+\frac{y}{z} \right)\left(7s-4\overline{m}_Q^2 \right)\right.\nonumber\\
&&\left.+\frac{1}{3}\left(\frac{z}{y^2}+\frac{y}{z^2} \right)m_Q^2\left[ 7+5\overline{m}_Q^2\,\delta\left(s-\overline{m}_Q^2 \right)\right]-\frac{1}{3}(y+z)\left(4s-3\overline{m}_Q^2 \right)\right\} \nonumber\\
&&-\frac{g_s^2\langle\bar{q}q\rangle^2}{1944\pi^4}\int_{y_i}^{y_f}dy \int_{z_i}^{1-y}dz\,(1-y-z)\left\{  3\left(\frac{z}{y}+\frac{y}{z} \right)\left(2s-\overline{m}_Q^2 \right)\right. \nonumber\\
&&\left.+\left(\frac{z}{y^2}+\frac{y}{z^2} \right)m_Q^2\left[ 1+\overline{m}_Q^2\,\delta\left(s-\overline{m}_Q^2\right)\right]+2(y+z)\left[8s-3\overline{m}_Q^2 +\overline{m}_Q^4\, \delta\left(s-\overline{m}_Q^2\right)\right]\right\}\, ,
\end{eqnarray}

\begin{eqnarray}
\rho_7^{0}(s)&=&\frac{m_Q^3\langle\bar{q}q\rangle}{1536\pi^2}\langle\frac{\alpha_sGG}{\pi}\rangle\int_{y_i}^{y_f}dy \int_{z_i}^{1-y}dz \left(\frac{y}{z^3}+\frac{z}{y^3}+\frac{1}{y^2}+\frac{1}{z^2}\right)(1-y-z) \nonumber\\
&&\left( 1+\frac{2\overline{m}_Q^2}{T^2}\right)\delta\left(s-\overline{m}_Q^2\right)\nonumber\\
&&-\frac{3m_Q\langle\bar{q}q\rangle}{256\pi^2}\langle\frac{\alpha_sGG}{\pi}\rangle\int_{y_i}^{y_f}dy \int_{z_i}^{1-y}dz \left(\frac{y}{z^2}+\frac{z}{y^2}\right)(1-y-z)
\left\{1+\frac{2\overline{m}_Q^2}{3}\delta\left(s-\overline{m}_Q^2\right) \right\} \nonumber\\
&&-\frac{m_Q\langle\bar{q}q\rangle}{128\pi^2}\langle\frac{\alpha_sGG}{\pi}\rangle\int_{y_i}^{y_f}dy \int_{z_i}^{1-y}dz\left\{1+\frac{2\overline{m}_Q^2}{3}\delta\left(s-\overline{m}_Q^2\right) \right\} \nonumber\\
&&-\frac{m_Q\langle\bar{q}q\rangle}{512\pi^2}\langle\frac{\alpha_sGG}{\pi}\rangle\int_{y_i}^{y_f}dy \left\{1+\frac{2\widetilde{m}_Q^2}{3}\delta\left(s-\widetilde{m}_Q^2\right) \right\}  \, ,
\end{eqnarray}

\begin{eqnarray}
\rho_7^{8}(s)&=&\frac{m_Q^3\langle\bar{q}q\rangle}{432\pi^2}\langle\frac{\alpha_sGG}{\pi}\rangle\int_{y_i}^{y_f}dy \int_{z_i}^{1-y}dz \left(\frac{y}{z^3}+\frac{z}{y^3}+\frac{1}{y^2}+\frac{1}{z^2}\right)(1-y-z) \nonumber\\
&&\left( 1+\frac{2\overline{m}_Q^2}{T^2}\right)\delta\left(s-\overline{m}_Q^2\right)\nonumber\\
&&-\frac{m_Q\langle\bar{q}q\rangle}{24\pi^2}\langle\frac{\alpha_sGG}{\pi}\rangle\int_{y_i}^{y_f}dy \int_{z_i}^{1-y}dz \left(\frac{y}{z^2}+\frac{z}{y^2}\right)(1-y-z)
\left\{1+\frac{2\overline{m}_Q^2}{3}\delta\left(s-\overline{m}_Q^2\right) \right\} \nonumber\\
&&+\frac{m_Q\langle\bar{q}q\rangle}{288\pi^2}\langle\frac{\alpha_sGG}{\pi}\rangle\int_{y_i}^{y_f}dy \int_{z_i}^{1-y}dz\left\{1+\frac{2\overline{m}_Q^2}{3}\delta\left(s-\overline{m}_Q^2\right) \right\} \nonumber\\
&&+t\frac{m_Q\langle\bar{q}q\rangle}{144\pi^2}\langle\frac{\alpha_sGG}{\pi}\rangle\int_{y_i}^{y_f}dy \int_{z_i}^{1-y}dz\left\{1-7\left(\frac{1}{y}+\frac{1}{z}\right)\frac{1-y-z}{4}\right\}\left\{1+\frac{2\overline{m}_Q^2}{3}\delta\left(s-\overline{m}_Q^2\right) \right\} \nonumber\\
&&-\frac{m_Q\langle\bar{q}q\rangle}{144\pi^2}\langle\frac{\alpha_sGG}{\pi}\rangle\int_{y_i}^{y_f}dy \left\{1+\frac{2\widetilde{m}_Q^2}{3}\delta\left(s-\widetilde{m}_Q^2\right) \right\}  \, ,
\end{eqnarray}

\begin{eqnarray}
\rho_8^{0}(s)&=&-\frac{m_Q^2\langle\bar{q}q\rangle\langle\bar{q}g_s\sigma Gq\rangle}{32\pi^2}\int_0^1 dy \left(1+\frac{\widetilde{m}_Q^2}{T^2} \right)\delta\left(s-\widetilde{m}_Q^2\right)\nonumber\\
&&+\frac{m_Q^2\langle\bar{q}q\rangle\langle\bar{q}g_s\sigma Gq\rangle}{64\pi^2}\int_0^1 dy \left( \frac{1}{y}+\frac{1}{1-y} \right)\delta\left(s-\widetilde{m}_Q^2\right)   \, ,
\end{eqnarray}

\begin{eqnarray}
\rho_8^{8}(s)&=&-\frac{m_Q^2\langle\bar{q}q\rangle\langle\bar{q}g_s\sigma Gq\rangle}{9\pi^2}\int_0^1 dy \left(1+\frac{\widetilde{m}_Q^2}{T^2} \right)\delta\left(s-\widetilde{m}_Q^2\right)\nonumber\\
&&-\frac{m_Q^2\langle\bar{q}q\rangle\langle\bar{q}g_s\sigma Gq\rangle}{144\pi^2}\int_0^1 dy \left( \frac{1}{y}+\frac{1}{1-y} \right)\delta\left(s-\widetilde{m}_Q^2\right)\nonumber\\
&&-t\frac{\langle\bar{q}q\rangle\langle\bar{q}g_s\sigma Gq\rangle}{144\pi^2}\int_{y_i}^{y_f} dy \left\{1+\frac{2\widetilde{m}_Q^2}{3}\delta\left(s-\widetilde{m}_Q^2\right) \right\}  \, ,
\end{eqnarray}

\begin{eqnarray}
\rho_{10}^{0}(s)&=&\frac{m_Q^2\langle\bar{q}g_s\sigma Gq\rangle^2}{256\pi^2T^6}\int_0^1 dy\, \widetilde{m}_Q^4\,\delta \left( s-\widetilde{m}_Q^2\right)\nonumber\\
&&-\frac{m_Q^4\langle\bar{q}q\rangle^2}{288T^4}\langle\frac{\alpha_sGG}{\pi}\rangle\int_0^1 dy  \left\{ \frac{1}{y^3}+\frac{1}{(1-y)^3}\right\} \delta\left( s-\widetilde{m}_Q^2\right)\nonumber\\
&&+\frac{m_Q^2\langle\bar{q}q\rangle^2}{96T^2}\langle\frac{\alpha_sGG}{\pi}\rangle\int_0^1 dy  \left\{ \frac{1}{y^2}+\frac{1}{(1-y)^2}\right\} \delta\left( s-\widetilde{m}_Q^2\right)\nonumber\\
&&-\frac{m_Q^2\langle\bar{q}g_s\sigma Gq\rangle^2}{256\pi^2T^4}\int_0^1 dy \left( \frac{1}{y}+\frac{1}{1-y}\right)\widetilde{m}_Q^2 \, \delta \left( s-\widetilde{m}_Q^2\right) \nonumber\\
&&+t\frac{\langle\bar{q}g_s\sigma Gq\rangle^2}{20736\pi^2}\int_0^1 dy \left(1+\frac{2\widetilde{m}_Q^2}{T^2}  \right)\delta \left( s-\widetilde{m}_Q^2\right)\nonumber\\
&&+\frac{m_Q^2\langle\bar{q}q\rangle^2}{288T^6}\langle\frac{\alpha_sGG}{\pi}\rangle\int_0^1 dy \, \widetilde{m}_Q^4 \, \delta\left( s-\widetilde{m}_Q^2\right) \, ,
\end{eqnarray}

\begin{eqnarray}
\rho_{10}^{8}(s)&=&\frac{m_Q^2\langle\bar{q}g_s\sigma Gq\rangle^2}{72\pi^2T^6}\int_0^1 dy \,\widetilde{m}_Q^4\,\delta \left( s-\widetilde{m}_Q^2\right)\nonumber\\
&&-\frac{m_Q^4\langle\bar{q}q\rangle^2}{81T^4}\langle\frac{\alpha_sGG}{\pi}\rangle\int_0^1 dy  \left\{ \frac{1}{y^3}+\frac{1}{(1-y)^3}\right\} \delta\left( s-\widetilde{m}_Q^2\right)\nonumber\\
&&+\frac{m_Q^2\langle\bar{q}q\rangle^2}{27T^2}\langle\frac{\alpha_sGG}{\pi}\rangle\int_0^1 dy  \left\{ \frac{1}{y^2}+\frac{1}{(1-y)^2}\right\} \delta\left( s-\widetilde{m}_Q^2\right)\nonumber\\
&&+7t\frac{\langle\bar{q}q\rangle^2}{1296}\langle\frac{\alpha_sGG}{\pi}\rangle\int_0^1 dy  \left( 1+\frac{2\widetilde{m}_Q^2}{T^2}\right) \delta\left( s-\widetilde{m}_Q^2\right)\nonumber\\
&&+\frac{m_Q^2\langle\bar{q}g_s\sigma Gq\rangle^2}{576\pi^2T^4}\int_0^1 dy \left( \frac{1}{y}+\frac{1}{1-y}\right)\widetilde{m}_Q^2 \, \delta \left( s-\widetilde{m}_Q^2\right)\nonumber\\
&&+t\frac{\langle\bar{q}g_s\sigma Gq\rangle^2}{864\pi^2}\int_0^1 dy \left(1+\frac{3\widetilde{m}_Q^2}{2T^2}+\frac{\widetilde{m}_Q^4}{T^4} \right)\delta \left( s-\widetilde{m}_Q^2\right)\nonumber\\
&&+t\frac{\langle\bar{q}g_s\sigma Gq\rangle^2}{5832\pi^2}\int_0^1 dy \left(1+\frac{2\widetilde{m}_Q^2}{T^2}  \right)\delta \left( s-\widetilde{m}_Q^2\right)\nonumber\\
&&+\frac{m_Q^2\langle\bar{q}q\rangle^2}{81T^6}\langle\frac{\alpha_sGG}{\pi}\rangle\int_0^1 dy \, \widetilde{m}_Q^4 \, \delta\left( s-\widetilde{m}_Q^2\right) \, ,
\end{eqnarray}
the subscripts  $0$, $3$, $4$, $5$, $6$, $7$, $8$, $10$ denote the dimensions of the  vacuum condensates, the superscripts  $0$, $8$ denote
the 0-0 type and 8-8 type interpolating currents respectively; $y_{f}=\frac{1+\sqrt{1-4m_Q^2/s}}{2}$,
$y_{i}=\frac{1-\sqrt{1-4m_Q^2/s}}{2}$, $z_{i}=\frac{ym_Q^2}{y s -m_Q^2}$, $\overline{m}_Q^2=\frac{(y+z)m_Q^2}{yz}$,
$ \widetilde{m}_Q^2=\frac{m_Q^2}{y(1-y)}$, $\int_{y_i}^{y_f}dy \to \int_{0}^{1}dy$, $\int_{z_i}^{1-y}dz \to \int_{0}^{1-y}dz$ when the $\delta$ functions $\delta\left(s-\overline{m}_Q^2\right)$ and $\delta\left(s-\widetilde{m}_Q^2\right)$ appear.

 In this article, we carry out the
operator product expansion to the vacuum condensates  up to dimension-10, and
assume  vacuum saturation for the  higher dimensional  vacuum condensates.
The condensates $\langle \frac{\alpha_s}{\pi}GG\rangle$, $\langle \bar{q}q\rangle\langle \frac{\alpha_s}{\pi}GG\rangle$,
$\langle \bar{q}q\rangle^2\langle \frac{\alpha_s}{\pi}GG\rangle$, $\langle \bar{q} g_s \sigma Gq\rangle^2$ and $g_s^2\langle \bar{q}q\rangle^2$ are the vacuum expectations
of the operators of the order
$\mathcal{O}(\alpha_s)$.  The four-quark condensate $g_s^2\langle \bar{q}q\rangle^2$ comes from the terms
$\langle \bar{q}\gamma_\mu t^a q g_s D_\eta G^a_{\lambda\tau}\rangle$, $\langle\bar{q}_jD^{\dagger}_{\mu}D^{\dagger}_{\nu}D^{\dagger}_{\alpha}q_i\rangle$  and
$\langle\bar{q}_jD_{\mu}D_{\nu}D_{\alpha}q_i\rangle$, rather than comes from the perturbative corrections of $\langle \bar{q}q\rangle^2$.
 The condensates $\langle g_s^3 GGG\rangle$, $\langle \frac{\alpha_s GG}{\pi}\rangle^2$,
 $\langle \frac{\alpha_s GG}{\pi}\rangle\langle \bar{q} g_s \sigma Gq\rangle$ have the dimensions 6, 8, 9 respectively,  but they are   the vacuum expectations
of the operators of the order    $\mathcal{O}( \alpha_s^{3/2})$, $\mathcal{O}(\alpha_s^2)$, $\mathcal{O}( \alpha_s^{3/2})$ respectively, and discarded.  We take
the truncations $n\leq 10$ and $k\leq 1$ in a consistent way,
the operators of the orders $\mathcal{O}( \alpha_s^{k})$ with $k> 1$ are  discarded. Furthermore,  the values of the  condensates $\langle g_s^3 GGG\rangle$, $\langle \frac{\alpha_s GG}{\pi}\rangle^2$,
 $\langle \frac{\alpha_s GG}{\pi}\rangle\langle \bar{q} g_s \sigma Gq\rangle$   are very small, and they can be  neglected safely.

 We differentiate   Eq.(23) with respect to  $\frac{1}{T^2}$,  eliminate the
 pole residues $\lambda_{X/Z}$, and obtain the QCD sum rules for
 the masses,
 \begin{eqnarray}
 M^2_{X/Z}= \frac{\int_{4m_Q^2}^{s_0} ds\frac{d}{d \left(-1/T^2\right)}\rho^{0/8}(s)e^{-\frac{s}{T^2}}}{\int_{4m_Q^2}^{s_0} ds \rho^{0/8}(s)e^{-\frac{s}{T^2}}}\, .
\end{eqnarray}

\section{Numerical results and discussions}
The input parameters are taken to be the standard values $\langle
\bar{q}q \rangle=-(0.24\pm 0.01\, \rm{GeV})^3$,   $\langle
\bar{q}g_s\sigma G q \rangle=m_0^2\langle \bar{q}q \rangle$,
$m_0^2=(0.8 \pm 0.1)\,\rm{GeV}^2$, $\langle \frac{\alpha_s
GG}{\pi}\rangle=(0.33\,\rm{GeV})^4 $    at the energy scale  $\mu=1\, \rm{GeV}$
\cite{SVZ79,Reinders85,Ioffe2005}.
The quark condensate and mixed quark condensate evolve with the   renormalization group equation, $\langle\bar{q}q \rangle(\mu)=\langle\bar{q}q \rangle(Q)\left[\frac{\alpha_{s}(Q)}{\alpha_{s}(\mu)}\right]^{\frac{4}{9}}$ and $\langle\bar{q}g_s \sigma Gq \rangle(\mu)=\langle\bar{q}g_s \sigma Gq \rangle(Q)\left[\frac{\alpha_{s}(Q)}{\alpha_{s}(\mu)}\right]^{\frac{2}{27}}$.

In the article, we take the $\overline{MS}$ masses $m_{c}(m_c)=(1.275\pm0.025)\,\rm{GeV}$ and $m_{b}(m_b)=(4.18\pm0.03)\,\rm{GeV}$
 from the Particle Data Group \cite{PDG}, and take into account
the energy-scale dependence of  the $\overline{MS}$ masses  from the renormalization group equation,
\begin{eqnarray}
m_c(\mu)&=&m_c(m_c)\left[\frac{\alpha_{s}(\mu)}{\alpha_{s}(m_c)}\right]^{\frac{12}{25}} \, ,\nonumber\\
m_b(\mu)&=&m_b(m_b)\left[\frac{\alpha_{s}(\mu)}{\alpha_{s}(m_b)}\right]^{\frac{12}{23}} \, ,\nonumber\\
\alpha_s(\mu)&=&\frac{1}{b_0t}\left[1-\frac{b_1}{b_0^2}\frac{\log t}{t} +\frac{b_1^2(\log^2{t}-\log{t}-1)+b_0b_2}{b_0^4t^2}\right]\, ,
\end{eqnarray}
  where $t=\log \frac{\mu^2}{\Lambda^2}$, $b_0=\frac{33-2n_f}{12\pi}$, $b_1=\frac{153-19n_f}{24\pi^2}$, $b_2=\frac{2857-\frac{5033}{9}n_f+\frac{325}{27}n_f^2}{128\pi^3}$,  $\Lambda=213\,\rm{MeV}$, $296\,\rm{MeV}$  and  $339\,\rm{MeV}$ for the flavors  $n_f=5$, $4$ and $3$, respectively  \cite{PDG}.

We  tentatively take the threshold parameters of the axial-vector molecular states $X(3872)$ (or $Z_c(3900)$) and $Z_b(10610)$   as $s_0=(18.5-20.5)\,\rm{GeV}^2$ and $(122-126)\,\rm{GeV}^2$  respectively
 to avoid the contaminations of the higher  resonances and continuum states, and search for the optimal values to satisfy the two criteria (pole dominance and convergence of the operator product
expansion) of the QCD sum rules. In this article, we  assume  that the energy gap between the ground states and
 the first radial excited states is about $(0.4-0.6)\,\rm{GeV}$, just like that of the conventional mesons.

The correlation functions $\Pi(p)$ can be written as
\begin{eqnarray}
\Pi(p)&=&\sum_n C_n(p^2,\mu)\langle{\mathcal{O}}_n(\mu)\rangle=\int_{4m^2_Q(\mu)}^\infty ds \frac{\rho_{QCD}(s,\mu)}{s-p^2} \nonumber\\
&=&\int_{4m^2_Q(\mu)}^{s_0} ds \frac{\rho_{QCD}(s,\mu)}{s-p^2}+\int_{s_0}^\infty ds \frac{\rho_{QCD}(s,\mu)}{s-p^2} \, ,
\end{eqnarray}
at the QCD side, where the $C_n(p^2,\mu)$ are the Wilson coefficients and the $\langle{\mathcal{O}}_n(\mu)\rangle$ are the vacuum condensates of dimension-$n$.
The short-distance contributions at  $p^2>\mu^2$ are included in the coefficients
$C_n(p^2,\mu)$, the long-distance contributions at $p^2<\mu^2$ are absorbed into the vacuum condensates  $\langle{\mathcal{O}}_n(\mu)\rangle$.
If $\mu\gg \Lambda_{QCD}$, the Wilson coefficients $C_n(p^2,\mu)$ depend only on short-distance dynamics,
 while the long-distance effects are
taken into account by the vacuum condensates $\langle{\mathcal{O}}_n(\mu)\rangle$.

The correlation functions $\Pi(p)$ are scale independent,
\begin{eqnarray}
\frac{d}{d\mu}\Pi(p)&=&0\, ,
\end{eqnarray}
which does not mean
\begin{eqnarray}
\frac{d}{d\mu}\int_{4m^2_Q(\mu)}^{s_0} ds \frac{\rho_{QCD}(s,\mu)}{s-p^2}\rightarrow 0 \, ,
\end{eqnarray}
 in the  present case due to the following two reasons:\\
$\bullet$ Perturbative corrections are neglected, the higher dimensional vacuum condensates are factorized into lower dimensional ones therefore  the energy scale dependence of the higher dimensional vacuum condensates is modified;\\
$\bullet$ Truncations $s_0$ set in, the correlation between the threshold $4m^2_Q(\mu)$ and continuum threshold $s_0$ is unknown,  the quark-hadron duality is an assumption.

We perform the Borel transform with respect to the variable $P^2=-p^2$  at the QCD side and obtain the result,
\begin{eqnarray}
\int_{4m^2_Q(\mu)}^{s_0} ds \frac{\rho_{QCD}(s,\mu)}{s-p^2}\rightarrow \int_{4m^2_Q(\mu)}^{s_0} ds \frac{\rho_{QCD}(s,\mu)}{T^2}\exp\left( -\frac{s}{T^2}\right) \, .
\end{eqnarray}
The QCD sum rules are  characterized by two energy scales $\mu^2$ and $T^2$.
The Borel parameters $T^2$ have to be small enough such that the contributions from the higher resonances  and
continuum states are damped sufficiently. On
the other hand, the Borel parameters $T^2$ must be large enough such that the higher-dimensional vacuum condensates
are suppressed sufficiently.

The heavy tetraquark system $Q\bar{Q}q^{\prime}\bar{q}$ could be described
by a double-well potential with two light quarks $q^{\prime}\bar{q}$ lying in the two wells respectively.
   In the heavy quark limit, the $c$ (and $b$) quark can be taken as a static well potential,
which binds the light quark $q^{\prime}$ to form a diquark in the color antitriplet channel or binds the light antiquark $\bar{q}$ to
form a meson in the color singlet channel (or a meson-like state in the color octet  channel).
Then the heavy tetraquark states  are characterized by the effective heavy quark masses ${\mathbb{M}}_Q$ (or constituent quark masses) and
the virtuality $V=\sqrt{M^2_{X/Y/Z}-(2{\mathbb{M}}_Q)^2}$ (or bound energy not as robust).
The effective masses  ${\mathbb{M}}_Q$ have uncertainties, the optimal values in the diquark-antidiquark systems are not necessary the ideal values in the
 meson-meson systems.

Now the QCD sum rules have three typical energy scales $\mu^2$, $T^2$, $V^2$.
 It is natural to take the energy  scale,
 \begin{eqnarray}
 \mu^2&=&V^2={\mathcal{O}}(T^2)\, .
 \end{eqnarray}

In Figs.1-3, we will plot only  the lines for the 0-0 type molecular states for simplicity.
In Fig.1,  the masses are plotted   with variations of the  Borel parameters $T^2$ and energy scales $\mu$ with the threshold parameters
$s_0=19.5\,\rm{GeV}^2$ and $124\,\rm{GeV}^2$  for the 0-0 type hidden charmed and hidden bottom  molecular states, respectively. From the figure, we can see that the masses decrease monotonously with increase of the energy scales, just like that of the tetraquark states \cite{WangHuangTao,Wang1311,Wang1312,WangHuangTao1312}.
If the energy scale formula
$\mu=\sqrt{M^2_{X/Y/Z}-(2{\mathbb{M}}_Q)^2}$
 with the effective masses ${\mathbb{M}}_c=1.80\,\rm{GeV}$ and ${\mathbb{M}}_b=5.13\,\rm{GeV}$ is also an acceptable  choice  in the case of the hadronic molecular states,
 the energy scales $\mu=1.5\,\rm{GeV}$ and $2.7\,\rm{GeV}$ for the  hidden charmed and hidden bottom  molecular states respectively   should reproduce the experimental
 values of the masses of the $X(3872)$, $Z_c(3900)$ and $Z_b(10610)$.

 In calculations, we observe that the effective masses
 ${\mathbb{M}}_c=1.80\,\rm{GeV}$ and ${\mathbb{M}}_b=5.13\,\rm{GeV}$ are acceptable values (if the uncertainties of the QCD sum rules are taken into account) but not the optimal values to reproduce the experimental values of the masses of the
 $X(3872)$, $Z_c(3900)$,  $Z_c(4020)$, $Z_c(4025)$, $Y(4140)$, $Z_b(10610)$ and $Z_b(10650)$ consistently in the scenario of molecular states \cite{Wang1403}.
 The energy scales $\mu=1.3\,\rm{GeV}$ and $2.6\,\rm{GeV}$
are the optimal energy scales to  reproduce the experimental data $M_{X(3872)}=3.87\,\rm{GeV}$, $M_{Z_c(3900)}=3.90\,\rm{GeV}$, $M_{Z_b(10610)}=10.61\,\rm{GeV}$
 (also the experimental values of the masses of the $Z_c(4020)$, $Z_c(4025)$, $Y(4140)$ and $Z_b(10650)$ \cite{Wang1403}) approximately.  The modified values ${\mathbb{M}}_c=1.84\,\rm{GeV}$ and ${\mathbb{M}}_b=5.14\,\rm{GeV}$ work for the hadronic molecular states, and can be
used to update the QCD sum rules for the heavy molecular states \cite{Wang4140}.

\begin{figure}
\centering
\includegraphics[totalheight=6cm,width=7cm]{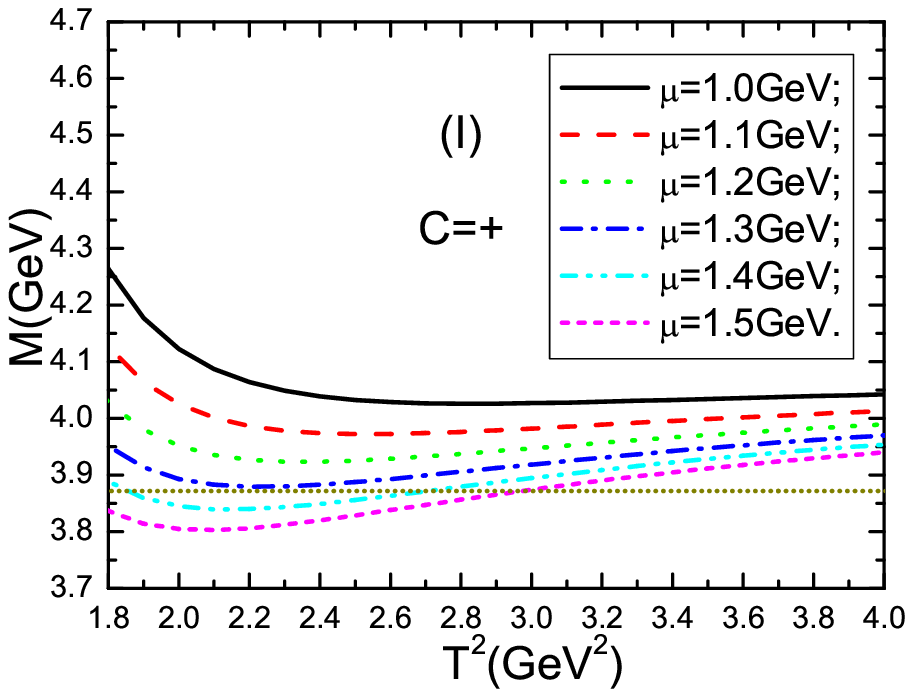}
\includegraphics[totalheight=6cm,width=7cm]{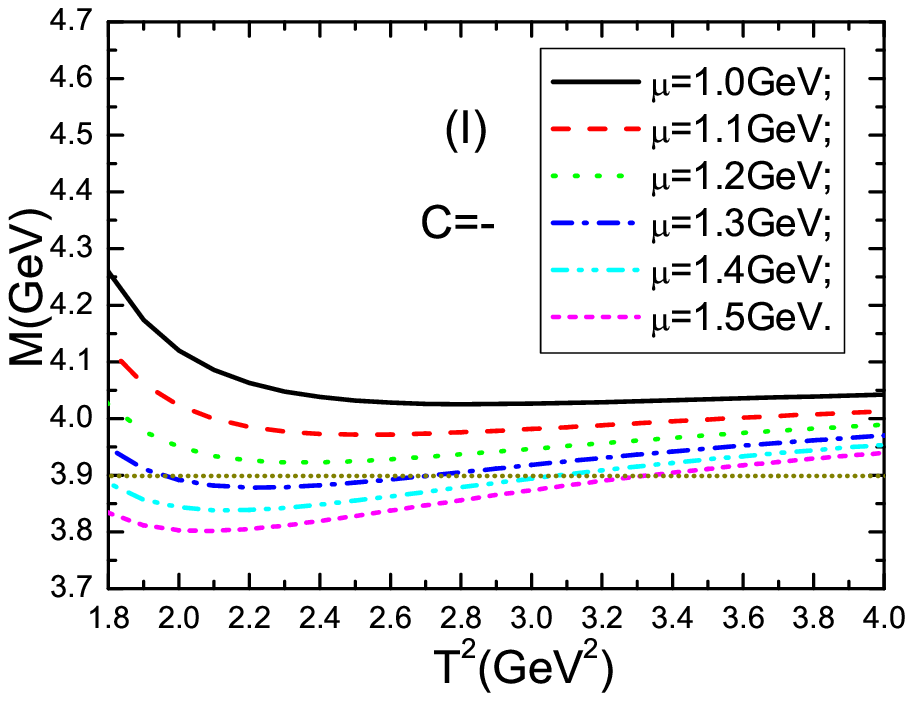}
\includegraphics[totalheight=6cm,width=7cm]{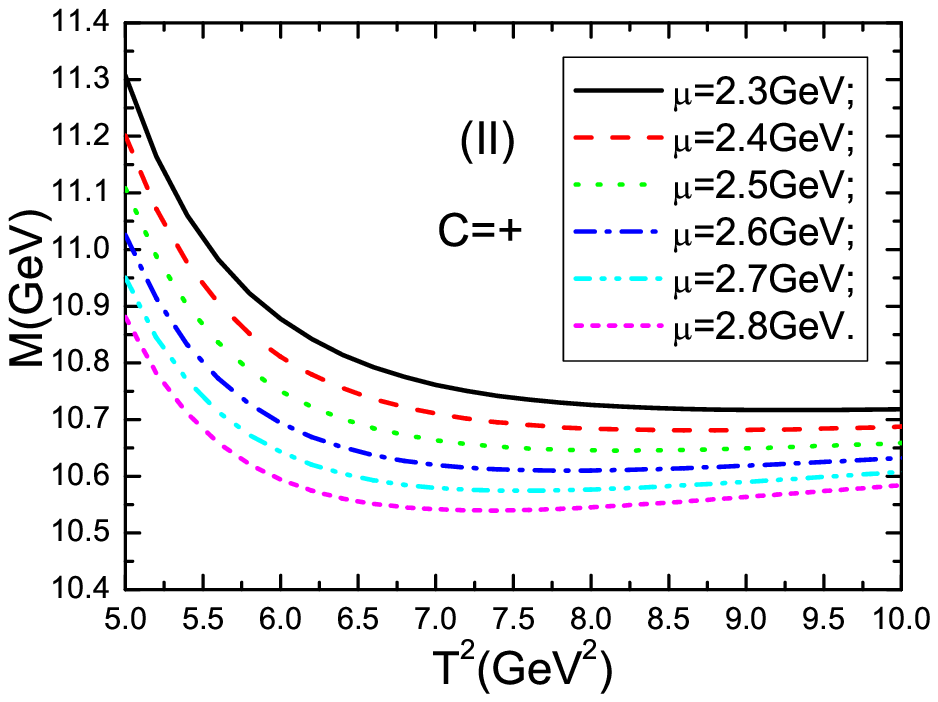}
\includegraphics[totalheight=6cm,width=7cm]{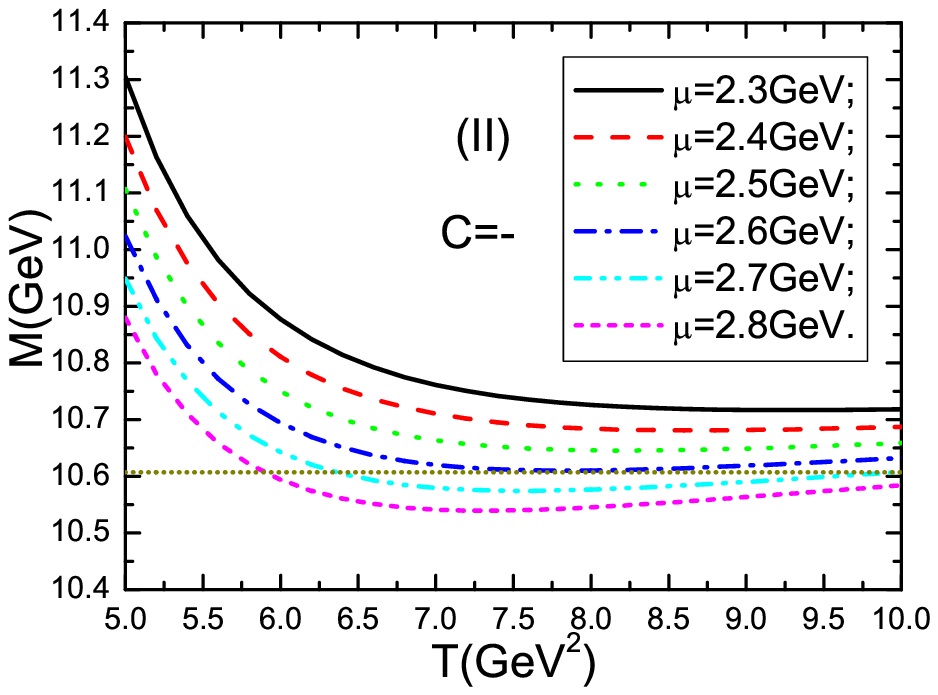}
  \caption{ The masses  with variations of the  Borel parameters $T^2$ and energy scales $\mu$, where the (I) and (II) denote the 0-0 type
  hidden charmed and hidden bottom molecular states respectively;  the horizontal lines denote the experimental values;
   the $C=\pm$ denote the charge conjugations. }
\end{figure}

\begin{figure}
\centering
\includegraphics[totalheight=6cm,width=7cm]{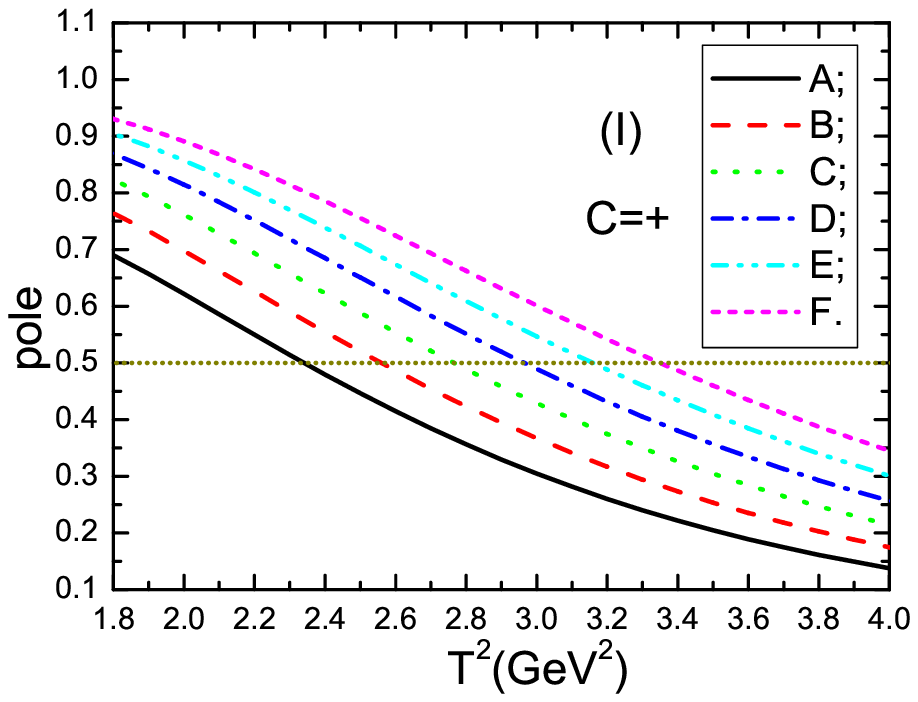}
\includegraphics[totalheight=6cm,width=7cm]{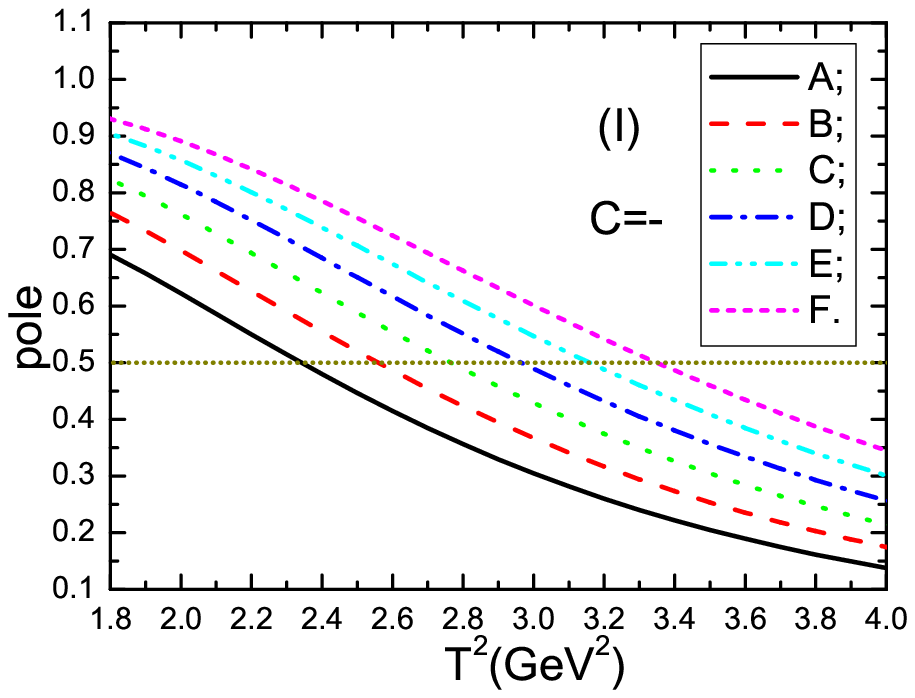}
\includegraphics[totalheight=6cm,width=7cm]{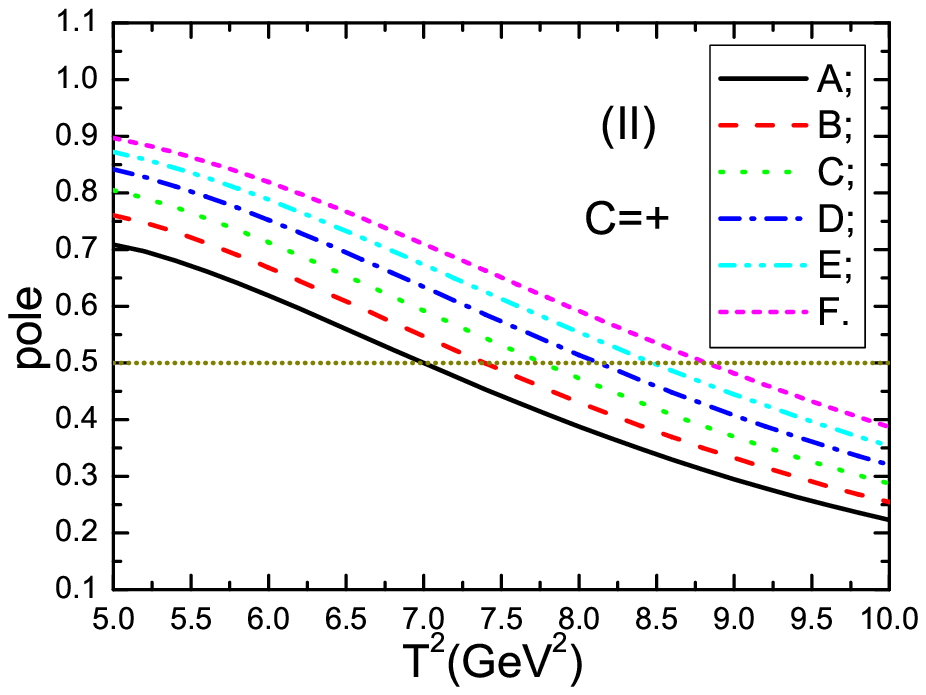}
\includegraphics[totalheight=6cm,width=7cm]{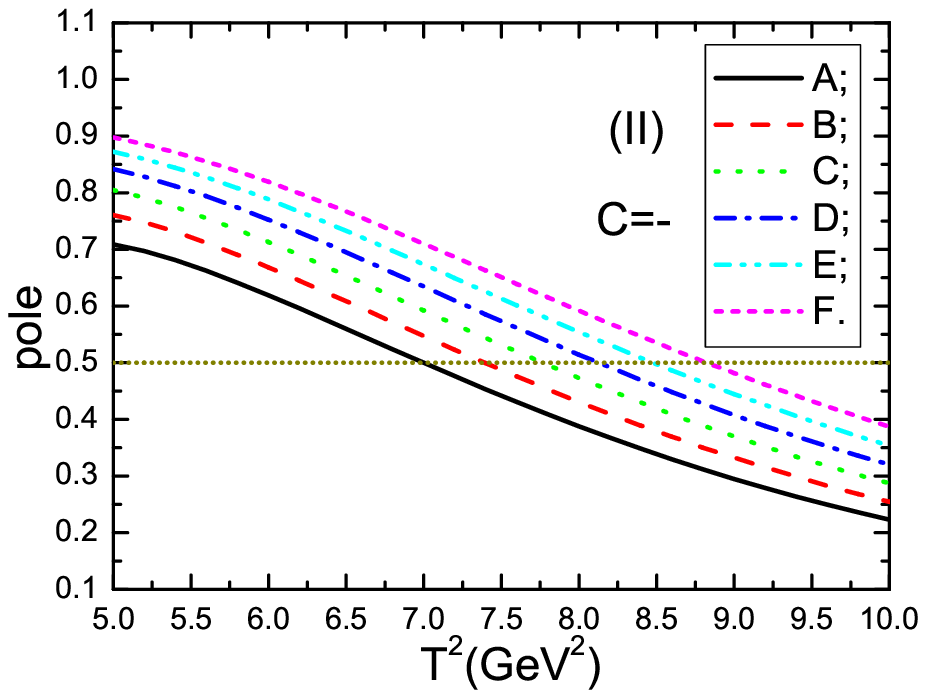}
  \caption{ The pole contributions  with variations of the  Borel parameters $T^2$ and threshold parameters $s_0$, where the  (I) and (II) denote the 0-0 type
  hidden charmed and hidden bottom molecular states respectively;
    the $A$, $B$, $C$, $D$, $E$, $F$ denote the threshold parameters $s_0=16.5$,  $17.5$, $18.5$, $19.5$, $20.5$, $21.5\,\rm{GeV}^2$ respectively
    for the hidden charmed molecular states, $s_0=118$,  $120$, $122$, $124$, $126$, $128\,\rm{GeV}^2$ respectively for the hidden bottom molecular states;
     the $C=\pm$ denote the charge conjugations. }
\end{figure}

In Fig.2,  the contributions of the pole terms are plotted with
variations of the threshold parameters $s_0$ and Borel parameters $T^2$ at the energy scales $\mu=1.3\,\rm{GeV}$ and $2.6\,\rm{GeV}$ for the 0-0 type hidden charmed and hidden bottom molecular states, respectively. In Fig.3,  the contributions of different terms in the
operator product expansion are plotted with variations of the Borel parameters  $T^2$ with  the  parameters
$s_0=19.5\,\rm{GeV}^2$, $\mu=1.3\,\rm{GeV}$ and $s_0=124\,\rm{GeV}^2$, $\mu=2.6\,\rm{GeV}$  for the 0-0 type hidden charmed and hidden bottom  molecular states, respectively. From the figures, we can choose the optimal Borel parameters and threshold parameters  to satisfy the two criteria  of the QCD sum rules.  The Borel parameters, continuum threshold parameters and the pole contributions are shown explicitly in Table 1.

\begin{table}
\begin{center}
\begin{tabular}{|c|c|c|c|c|c|c|c|}\hline\hline
$J^{PC}$                           &$T^2 (\rm{GeV}^2)$ &$s_0 (\rm{GeV}^2)$ &pole         & $M_{X/Z}(\rm{GeV})$     & $\lambda_{X/Z}(\rm{GeV}^5)$ \\ \hline
$1^{++}\,(\bar{u}c\bar{c}d)_{0-0}$ &$2.2-2.8$          &$19.5\pm1$         &$(49-80)\%$  & $3.89^{+0.09}_{-0.09}$  & $1.72^{+0.29}_{-0.25}\times10^{-2}$ \\ \hline
$1^{+-}\,(\bar{u}c\bar{c}d)_{0-0}$ &$2.2-2.8$          &$19.5\pm1$         &$(49-80)\%$  & $3.89^{+0.09}_{-0.09}$  & $1.72^{+0.29}_{-0.25}\times10^{-2}$ \\ \hline
$1^{++}\,(\bar{u}b\bar{b}d)_{0-0}$ &$7.2-8.0$          &$124\pm2$          &$(47-65)\%$  & $10.61^{+0.10}_{-0.09}$ & $1.13^{+0.17}_{-0.14}\times10^{-1}$ \\ \hline
$1^{+-}\,(\bar{u}b\bar{b}d)_{0-0}$ &$7.2-8.0$          &$124\pm2$          &$(47-65)\%$  & $10.61^{+0.10}_{-0.09}$ & $1.13^{+0.17}_{-0.14}\times10^{-1}$ \\ \hline
$1^{++}\,(\bar{u}c\bar{c}d)_{8-8}$ &$2.6-3.3$          &$22\pm1$           &$(51-80)\%$  & $4.08^{+0.10}_{-0.10}$  & $5.70^{+0.98}_{-0.81}\times10^{-2}$ \\ \hline
$1^{+-}\,(\bar{u}c\bar{c}d)_{8-8}$ &$2.6-3.3$          &$22\pm1$           &$(50-79)\%$  & $4.10^{+0.09}_{-0.10}$  & $5.75^{+0.97}_{-0.80}\times10^{-2}$ \\ \hline
$1^{++}\,(\bar{u}b\bar{b}d)_{8-8}$ &$7.4-8.2$          &$126\pm2$          &$(50-67)\%$  & $10.66^{+0.11}_{-0.08}$ & $2.63^{+0.37}_{-0.32}\times10^{-1}$ \\ \hline
$1^{+-}\,(\bar{u}b\bar{b}d)_{8-8}$ &$7.4-8.2$          &$126\pm2$          &$(50-67)\%$  & $10.66^{+0.11}_{-0.09}$ & $2.63^{+0.37}_{-0.31}\times10^{-1}$ \\ \hline
 \hline
\end{tabular}
\end{center}
\caption{ The Borel parameters, continuum threshold parameters, pole contributions, masses and pole residues of the 0-0 type and 8-8 type molecular states. }
\end{table}

 \begin{figure}
\centering
\includegraphics[totalheight=6cm,width=7cm]{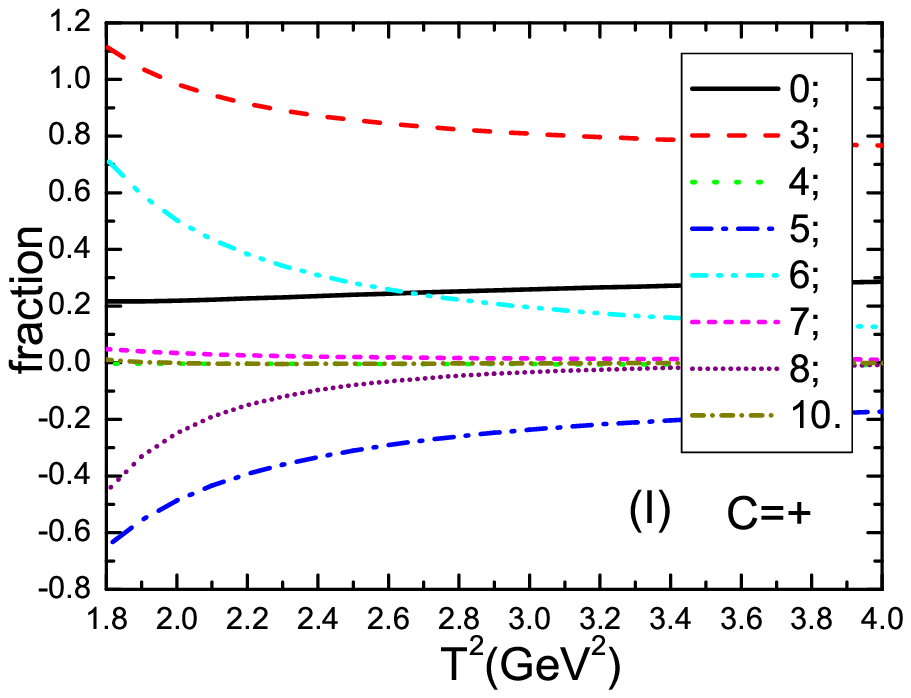}
\includegraphics[totalheight=6cm,width=7cm]{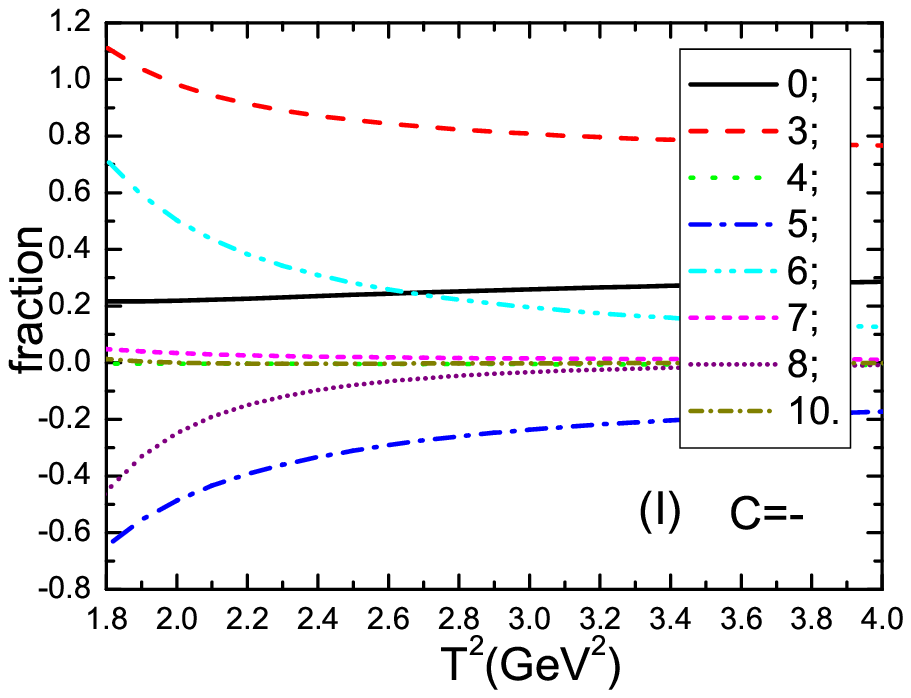}
\includegraphics[totalheight=6cm,width=7cm]{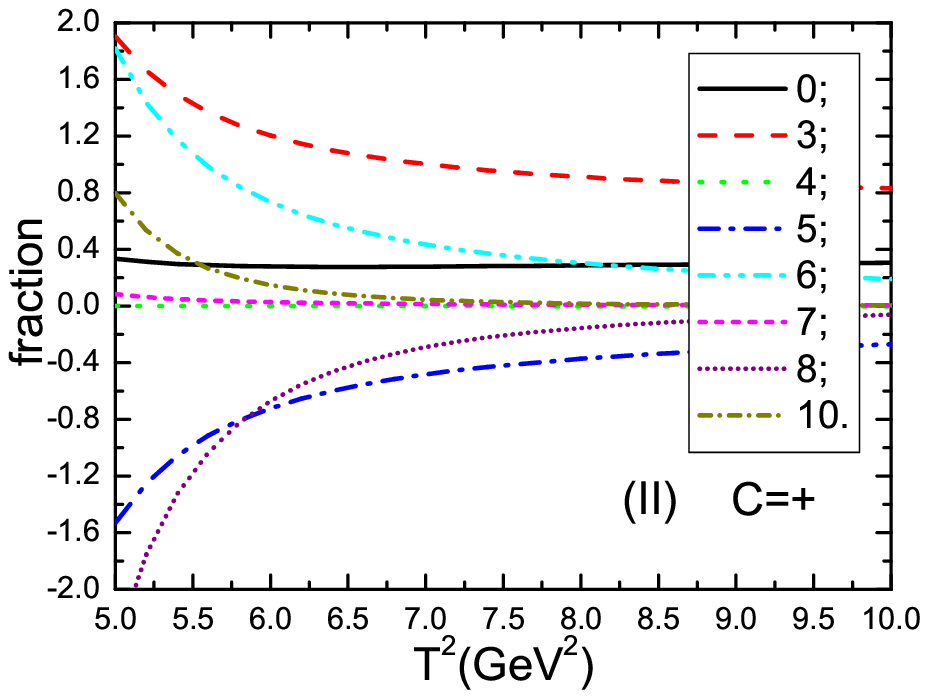}
\includegraphics[totalheight=6cm,width=7cm]{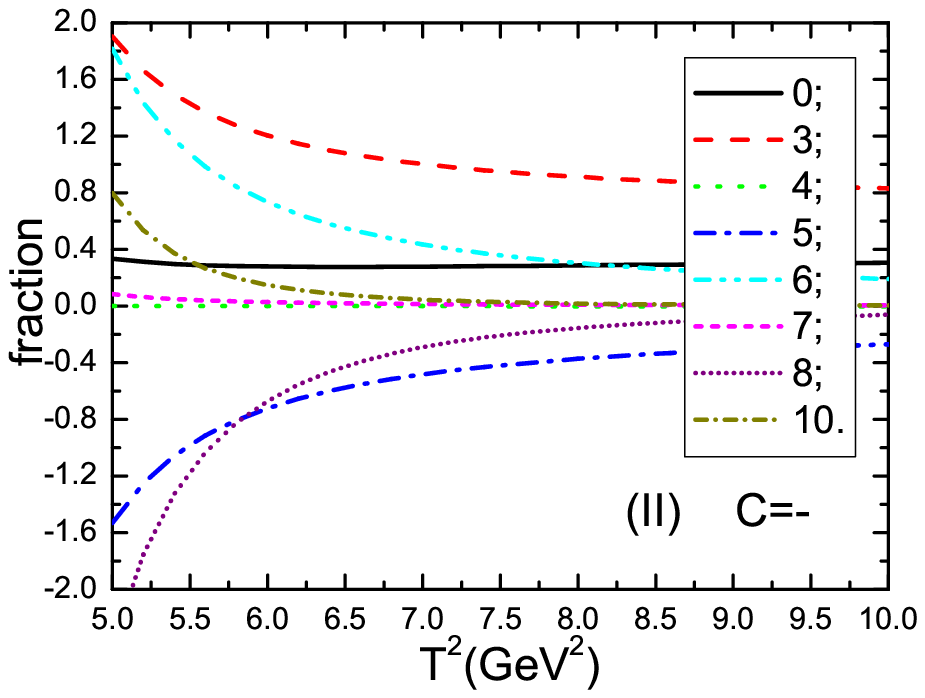}
  \caption{ The contributions of different terms in the operator product expansion  with variations of the  Borel parameters $T^2$,
  where the  (I) and (II) denote the 0-0 type  hidden charmed and hidden bottom molecular states respectively;
   the 0, 3, 4, 5, 6, 7, 8, 10 denote the dimensions of the vacuum condensates;  the $C=\pm$ denote the charge conjugations. }
\end{figure}

\begin{figure}
\centering
\includegraphics[totalheight=6cm,width=7cm]{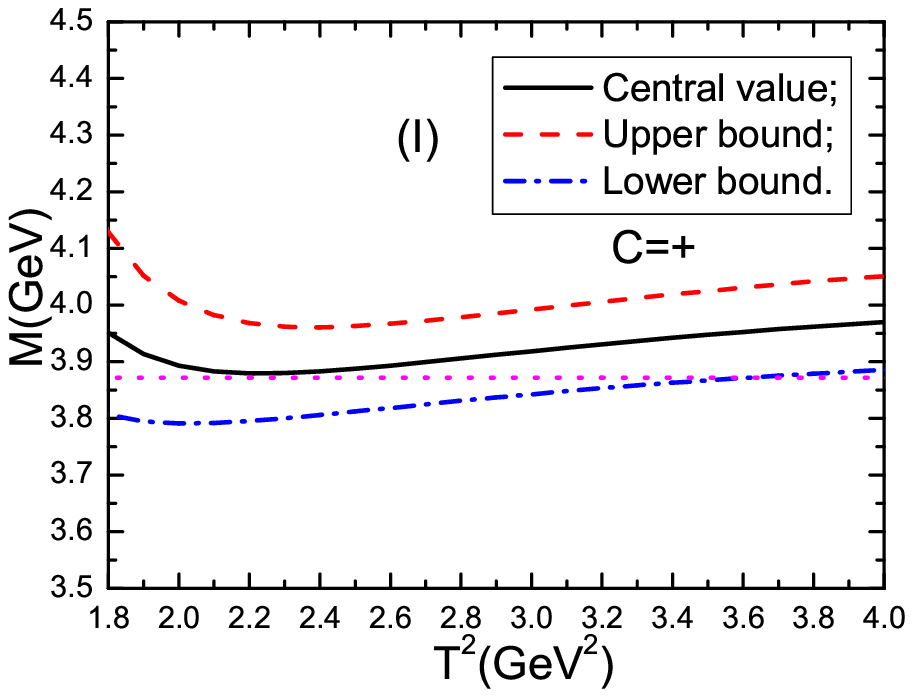}
\includegraphics[totalheight=6cm,width=7cm]{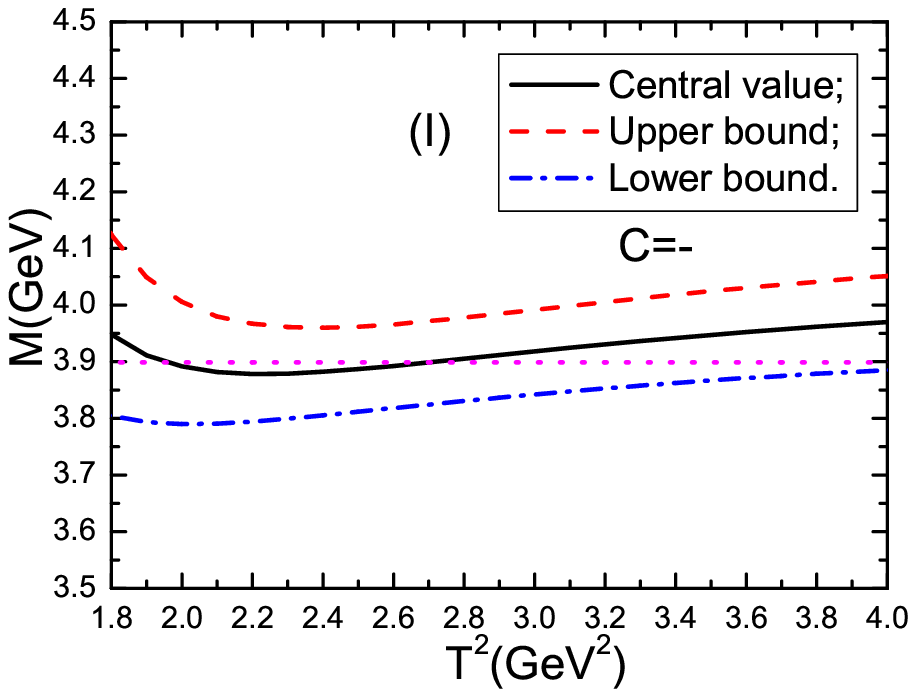}
\includegraphics[totalheight=6cm,width=7cm]{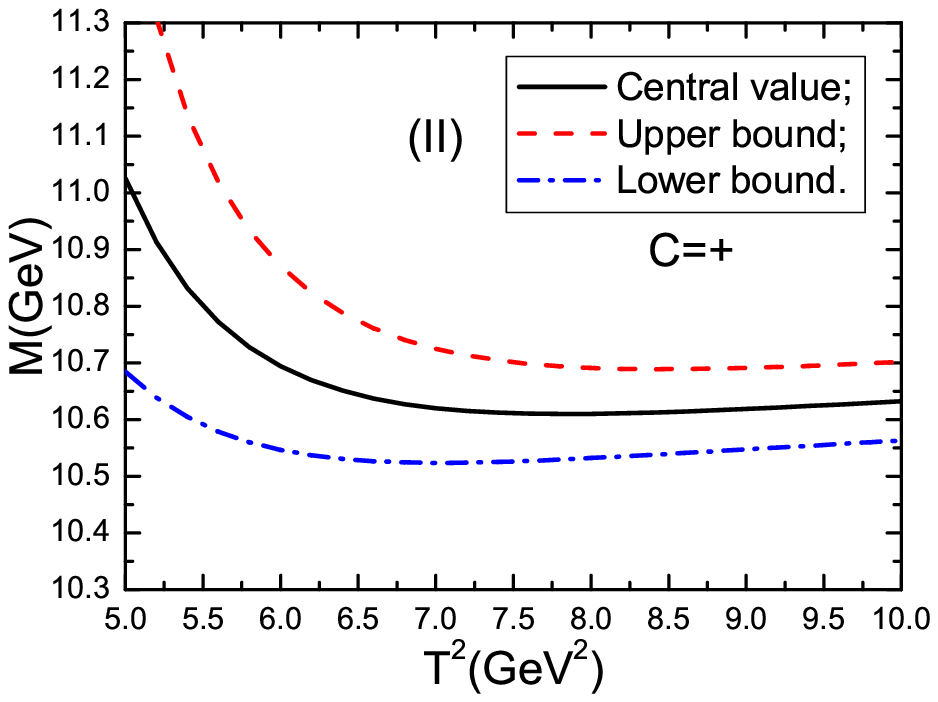}
\includegraphics[totalheight=6cm,width=7cm]{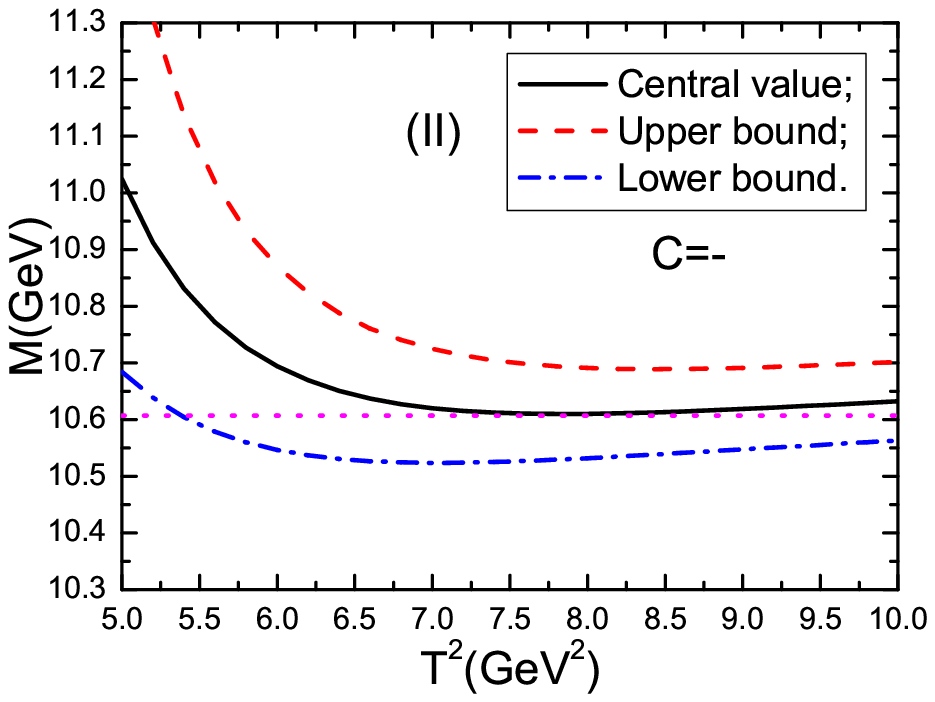}
  \caption{ The masses of the 0-0 type molecular states with variations of the  Borel parameters $T^2$,
  where the horizontal lines denote  the experimental values;
  the   (I) and (II) denote the  hidden charmed and hidden bottom molecular states, respectively;
  the $C=\pm$ denote the charge conjugations.}
\end{figure}

\begin{figure}
\centering
\includegraphics[totalheight=6cm,width=7cm]{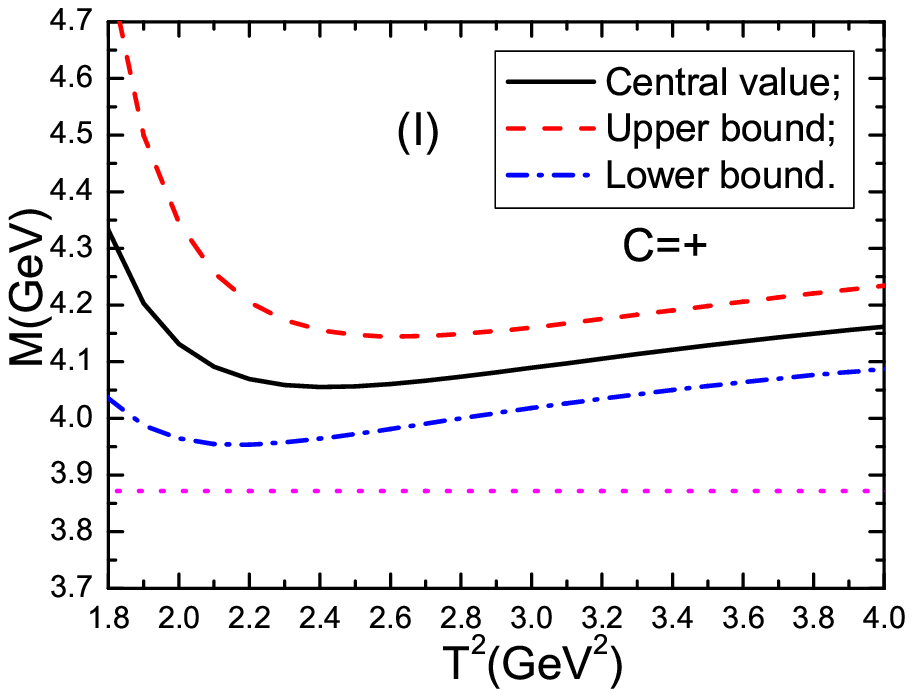}
\includegraphics[totalheight=6cm,width=7cm]{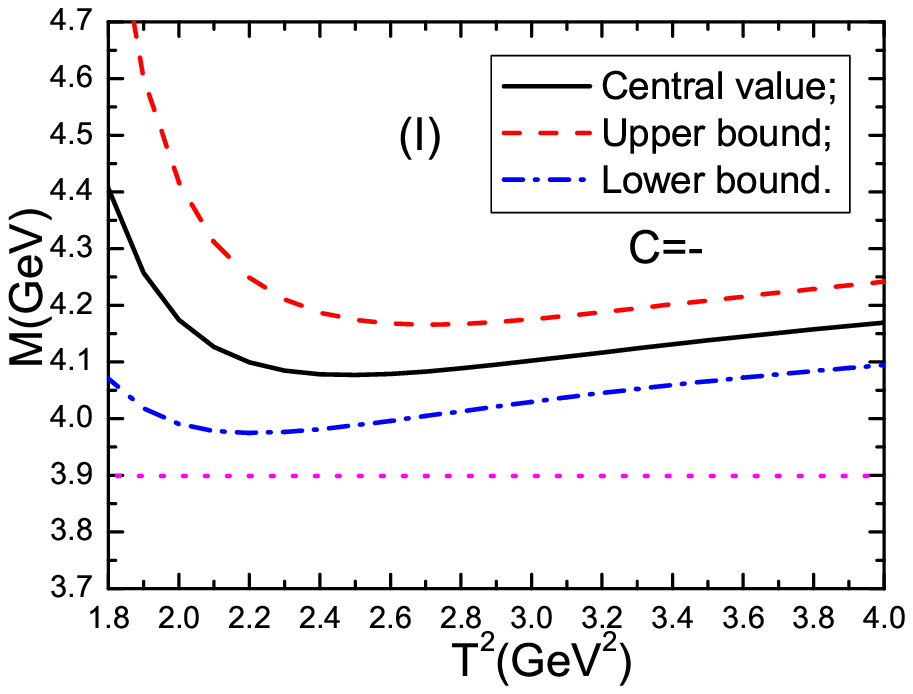}
\includegraphics[totalheight=6cm,width=7cm]{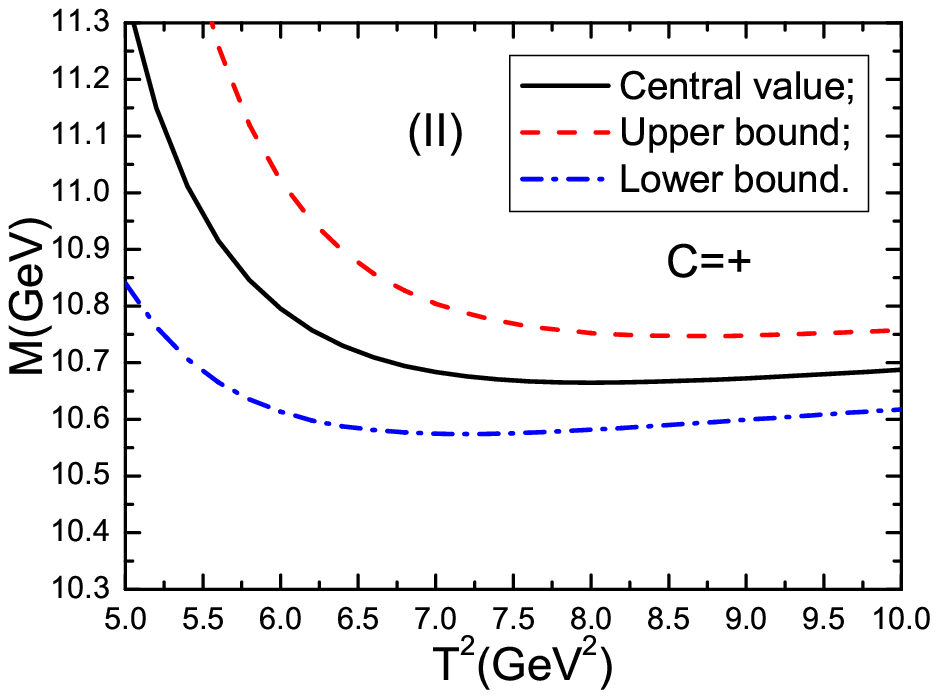}
\includegraphics[totalheight=6cm,width=7cm]{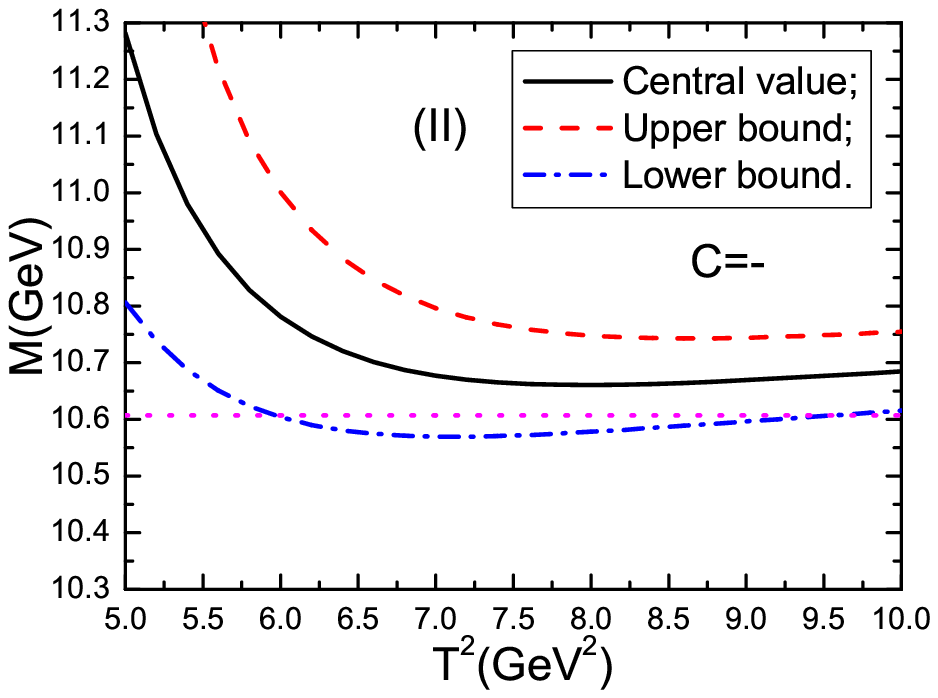}
  \caption{ The masses of the 8-8 type molecular states with variations of the  Borel parameters $T^2$,
  where the horizontal lines denote  the experimental values;
  the   (I) and (II) denote the hidden charmed and hidden bottom molecular states, respectively;
   the $C=\pm$ denote the charge conjugations.}
\end{figure}

We take into account all uncertainties of the input parameters,
and obtain the values of the masses and pole residues of
 the   molecular states, which are  shown in Table 1 and Figs.4-5.

 The masses of the 0-0 type molecular states $\bar{u}c\bar{c}d (1^{++})$, $\bar{u}c\bar{c}d (1^{+-})$ and $\bar{u}b\bar{b}d (1^{+-})$ are consistent with that of the $X(3872)$, $Z_c(3900)$ and $Z_b(10610)$ respectively within uncertainties,
 \begin{eqnarray}
 M_{\bar{u}c\bar{c}d,1^{++},(0-0)}&=&\left(3.89^{+0.09}_{-0.09}\right)\,{\rm{GeV}} \approx  M_{X(3872)}=(3871.68\pm 0.17 )\,\rm{MeV}\,(\rm{exp}) \cite{PDG}\, , \\
 M_{\bar{u}c\bar{c}d,1^{+-},(0-0)}&=&\left(3.89^{+0.09}_{-0.09}\right)\,{\rm{GeV}} \approx  M_{Z_c(3900)}=(3899.0\pm 3.6\pm 4.9)\,\rm{ MeV}\,(\rm{exp})\cite{BES3900} \, , \\
  M_{\bar{u}b\bar{b}d,1^{+-},(0-0)}&=&\left(10.61^{+0.10}_{-0.09}\right)\,{\rm{GeV}} \approx  M_{Z_b(10610)}=\left(10607.2\pm2.0\right)\,\rm{ MeV}\,(\rm{exp})\cite{Belle1110} \, .
 \end{eqnarray}
 The present predictions favor assigning the $X(3872)$, $Z_c(3900)$, $Z_b(10610)$ as the $S$-wave $D^*\bar{D}$, $D^*\bar{D}$ and $B^*\bar{B}$ molecular states, respectively, while our previous works
favor assigning the $X(3872)$, $Z_c(3900)$, $Z_b(10610)$ as the diquark-antidiquark type tetraquark  states \cite{WangHuangTao,WangHuangTao1312}.

Although the mass is a fundamental parameter in describing a hadron, a hadron cannot be identified  unambiguously by the mass alone, more  theoretical and experimental
works on the productions  and decays are still needed to identify the $X(3872)$, $Z_c(3900)$, $Z_b(10610)$. At the present time, it is still a open problem. From Table 1, we can see that the charge conjugation partners have
 almost degenerate masses, and the 8-8 type molecular states have larger masses than that of the 0-0 type molecular states. The present
 predictions can be confronted with the experimental data in the future at the BESIII, LHCb and Belle-II.

\section{Conclusion}
In this article, we  take the $X(3872)$, $Z_c(3900)$, $Z_b(10610)$ as the molecular states, construct both the color singlet-singlet type and color octet-octet type
currents to interpolate them, and calculate the  vacuum condensates up to dimension-10  in the operator product expansion.
Then we study  the axial-vector   hidden charmed and hidden bottom molecular states with the QCD sum rules, explore the energy scale dependence
in details  for the first time,
and use the energy scale formula
$\mu=\sqrt{M^2_{X/Y/Z}-(2{\mathbb{M}}_Q)^2}$
suggested in our previous works with the modified effective masses ${\mathbb{M}}_c=1.84\,\rm{GeV}$ and ${\mathbb{M}}_b=5.14\,\rm{GeV}$ to determine the
energy scales of the QCD spectral densities.  The energy scale formula
 works well for both the hidden charmed (or bottom) molecular states and tetraquark states.
 In the QCD sum rules for the hidden charmed (or bottom) tetraquark states and molecular states, the hadronic masses and pole residues
are sensitive to the heavy quark masses $m_Q$, the energy scale formula has outstanding advantage in determining  the $m_Q$.
 The numerical results  support assigning  the $X(3872)$, $Z_c(3900)$ and $Z_b(10610)$) as the 0-0 type  molecular states with $J^{PC}=1^{++}$, $1^{+-}$, $1^{+-}$, respectively;
while there are no candidates for the 8-8 type molecular states. The present predictions can be confronted with the experimental data in the future at the BESIII, LHCb
 and Belle-II.  More  theoretical and experimental
works on the productions  and decays are still needed to distinguish the molecule and tetraquark assignments, as a hadron cannot be identified  unambiguously by  the mass alone.
The  pole residues can be taken as   basic input parameters to study relevant processes of the $X(3872)$, $Z_c(3900)$ and
 $Z_b(10610)$ with the three-point QCD sum rules.

\section*{Acknowledgements}
This  work is supported by National Natural Science Foundation,
Grant Numbers 11375063, 11235005,  and the Fundamental Research Funds for the
Central Universities.

\end{document}